\documentclass{aa}
\usepackage{txfonts}
\usepackage{graphicx}
\usepackage{longtable}
\usepackage{lscape}
\newcommand{\ha}{H$\alpha$}
\newcommand{\hii}{H$_\mathrm{II}$ }
\newcommand{\e}{et al.\ }
\newcommand{\ngc}{NGC~6530}
\begin{document}

\title{A wide-area photometric and astrometric (Gaia DR2) study of the young
cluster NGC~6530
\thanks{Table~\ref{table-members} is only available in
electronic form at the CDS via anonymous ftp to cdsarc.u-strasbg.fr
(130.79.125.5) or via http://cdsweb.u-strasbg.fr/Abstract.html}
}

\date{Received date / Accepted date}

\author{F. Damiani\inst{1}
\and L. Prisinzano\inst{1}
\and G. Micela\inst{1} \and S. Sciortino\inst{1}
}
\institute{INAF - Osservatorio Astronomico di Palermo G.S.Vaiana,
Piazza del Parlamento 1, I-90134 Palermo, ITALY
}

\abstract{
NGC~6530 is a young cluster, with a complex morphology and
star-formation history. We present a statistical study of its
global properties, using a new, large list of candidate members down to
masses of $0.2-0.4 M_{\odot}$ and Gaia DR2 astrometry.
}{
We consider a larger sky region compared to previous studies, to
investigate the entire cluster until its periphery, including any diffuse
population all around the main cluster. We study the distribution of
extinction and age across the different regions, and obtain
constraints on the star-formation history. We also study the
dynamics of cluster members.
}{
Cluster membership is determined on the basis of literature X-ray data,
\ha\ emission, near-IR and UV excesses from the VPHAS+ and UKIDSS
photometric surveys and published near-IR catalogs, and Gaia DR2
astrometry; moreover, we use a method for photometric selection of M-type 
pre-main-sequence cluster members, which we recently developed and used for
other star-formation regions. The list of candidates includes nearly
{ 3700} stars, of which we estimate approximately { 2700}
genuine \ngc\ members.
}{
Using Gaia parallaxes, the cluster distance is found to be 1325~pc, with errors
of 0.5\% (statistical) and 8.5\% (systematic), in agreement with
previous determinations.
The cluster morphology and boundaries are established with great
confidence, from the agreement between the subsamples of members
selected using different criteria. There is no diffuse population of
members around the cluster, but there are minor condensations of true
members in addition to the two main groups in the cluster core and in
the Hourglass nebula. Two such subgroups are spatially associated with the
{ stars 7~Sgr (F2II-III) and HD~164536 (O7.5V).}
There is a definite pattern of
sequential star formation across the cluster, within an age range from
less than 0.5~Myr to $\sim 5$~Myr. Extinction is spatially
non-uniform, with part of the population still embedded or obscured by
thick dust. 
The precise Gaia proper motion data indicate that the \ngc\ parent cloud
collided
with the Galactic plane around 4~Myr ago, and we suggest that event
as the trigger of the bulk of star formation in \ngc. The internal cluster
dynamics is also partially resolved by the Gaia data, indicating
expansion of the main cluster population with respect to its center.
}{}

\keywords{Open clusters and associations: individual (NGC~6530)
-- Stars: pre-main-sequence -- Parallaxes -- Proper motions
-- X-rays: stars
}

\titlerunning{A photometric and astrometric study of NGC~6530}
\authorrunning{Damiani et al.}

\maketitle

\section{Introduction}
\label{intro}

\ngc\ is a young cluster, only a few Myr old, associated to the \hii
region M8 (Lagoon Nebula), which was studied using a variety of methods
in the past decades (see Tothill 2008 for a review). Its low-mass stars
are still found in their pre-main sequence (PMS) evolutionary phase, and
therefore they are bright X-ray sources
{ due to their active coronae (e.g., Feigelson and Montmerle 1999)},
show often \ha\ emission lines { and UV excess emission from
accretion of matter onto the star surface,}
and near-IR (NIR) emission { from circumstellar dusty disks.}
All these characteristics have
helped to detect a large population of { cumulatively $\sim 2500$}
candidate cluster
members down to sub-solar masses { (e.g., $\sim 1100$ members were
found using X-rays and NIR excesses by Damiani \e 2004, 2006a, and} Prisinzano
\e 2005, 2007; { 232 members found using Spitzer data by}
Kumar and Anandarao 2010; { 2427 candidate members found using deeper
Chandra X-ray data by} Kuhn \e 2013, 2014, 2015; { 235 members found
using photometric $u$ and \ha\ data by} Kalari
\e 2015), later confirmed by spectroscopic studies (Prisinzano \e 2012;
Prisinzano \e in preparation). The member stars exhibit a well-defined
spatial structure, with strong indications that some regions (M8-East,
Hourglass Nebula) are younger than the bulk of the population.
{ M8-East and the Hourglass Nebula are associated with the Lagoon
Nebula and are therefore assumed to lie at the same distance.}
A similar
complexity was also found from the analysis of the large-scale kinematics
of the ionized and neutral gas (Damiani \e 2017a) using several optical
emission and absorption lines.
{ Using X-ray membership and deep optical photometry,}
Prisinzano \e (2005) redetermined the cluster distance to be 1250~pc,
sensibly lower than previous values found in the literature
{ (1800~pc, Sung \e 2000, van den Ancker \e 1997).} The
average optical extinction is $A_V=1.085$~mag from Sung \e (2000).

The cluster member list obtained from all these studies cannot however
be considered complete, since each of the methods used suffers from some
bias, either inherent to its nature or dictated by the availability and
quality of observational data. For example, X-ray imaging data obtained
with the Chandra X-ray Observatory (Damiani \e 2004, Kuhn \e 2013)
consist of only two pointings with the ACIS-I CCD imaging spectrometer,
with a square field-of-view (FOV) of 17$^{\prime}$ on a side. The two
pointings are of very different exposure time (one of $\sim$~60~ks centered on
cluster core, the other of $\sim$~180~ks centered on the Hourglass nebula).
Inside the ACIS FOV, the limiting sensitivity is not uniform, degrading
rapidly to 1/4 of its on-axis value near the FOV borders, because of the
degradation of the PSF width with off-axis angle. On the other hand, the
relatively hard X-ray emission from PMS low-mass stars is relatively
insensitive to absorption, which becomes a distinct advantage in the
most obscured parts of the cluster. However, since X-ray luminosity in
this age range scales in proportion to stellar bolometric luminosity,
the lowest stellar masses are only detected in the deepest X-ray
observations.

Member selection methods based on \ha\ emission, NIR and UV excess of
PMS stars are by definition only selecting stars with strong lines from
disk-accretion phenomena - so-called Classical T~Tauri Stars (CTTS) -
{ since the accreted matter becomes hot enough to emit strongly in the UV
continuum and in the Balmer lines, or stars with}
massive circumstellar dusty disks { which emit in the NIR more
than the stellar photosphere} (Class~I/II stars).
However, these are
as a rule only one component (often a secondary one) of the total
population of even very young clusters, the rest of the members
belonging to the Weak-line T~Tauri Star (WTTS, or Class~III) population.

We have recently developed a method (Damiani 2018; Paper~I), which relies on the specific
properties of PMS M stars, to select candidate low-mass members of PMS
clusters using deep multi-band photometry, such as is becoming available
from the several wide area surveys in the optical and NIR (e.g.\ in the
optical: VPHAS+, Drew \e 2014; PanSTARRS, Chambers \e 2016; in the NIR:
UKIDSS, Lawrence \e 2007; VVV, Minniti \e 2011). The method was tested
on the Sco~OB1 star-forming complex, where approximately 4000 M-type
candidate members were found in Paper~I, and also in the Vela
Molecular Ridge region (Prisinzano \e 2018) and NGC~2264 cluster
(Venuti \e 2018). This method is unbiased with respect to
disk- and accretion-related excess emission (or is maybe slightly biased against
them, thus becoming complementary to those methods).
Since this method requires photometry in the optical $r$ and $i$ bands
{ (besides one of $g$, $J$, or $H$ bands)},
it becomes less powerful for extincted regions, so that again there is some
complementarity with X-ray imaging.
The limitation inherent to the method, of selecting exclusively M-type
members, still makes it very powerful, since these stars constitute the
majority of the population of any cluster.
We therefore examine here if any sizable population of M stars is
present in \ngc, and the constraints it adds to the cluster structure
and formation history.

Finally, the recently released astrometric data (DR2) from the Gaia mission
(Gaia Collaboration 2016, 2018) provide us with an additional method for
finding candidate
members of \ngc, on the basis of their parallax and proper motion. Since the
precision of both quantities decreases towards fainter magnitudes,
astrometric selection becomes increasingly ineffective towards the lowest
stellar masses. Both the astrometric and photometric data used here are
available over a wide sky area, thus enabling us to study the cluster
together with its surroundings.

{ The purpose of this work is to assemble a most comprehensive list of
candidate members of NGC~6530 using the various methods described above,
to use this dataset to study global properties of \ngc,
and to study to what extent those methods are complementary to one
another.}

This paper is structured as follows: in Sect.~\ref{obs} we describe the
observational data used. In Sect.~\ref{members} the various methods to
select candidate members are presented. Results are studied in
Sect.~\ref{results}, and Sect.~\ref{concl} provides a conclusive discussion.

\begin{figure}
\resizebox{\hsize}{!}{
\includegraphics[]{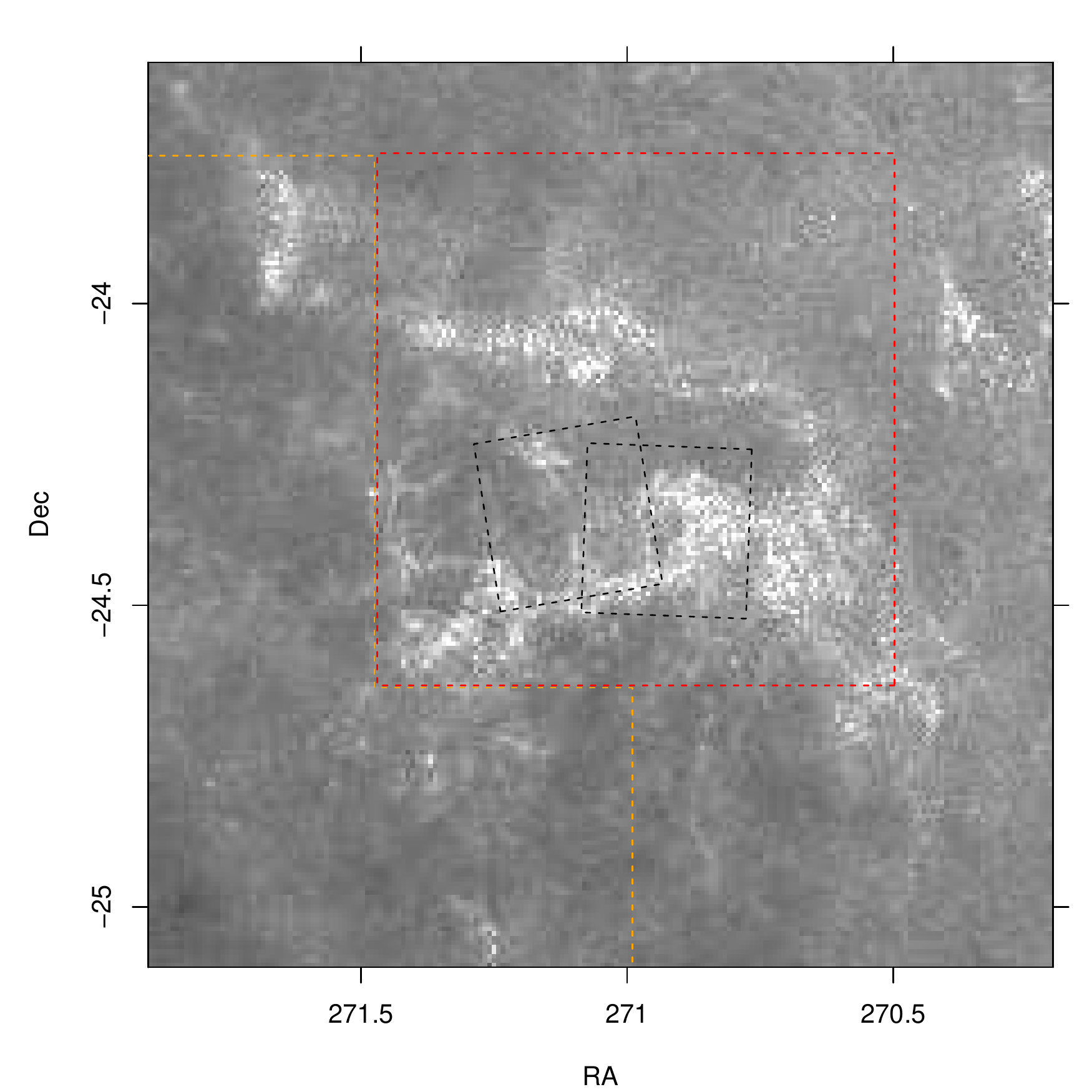}}
\caption{
Spatial density of all Gaia sources in the region studied.
Light-gray corresponds to the lowest source density. The two dashed
black squares are the Chandra ACIS-I FOVs.
The dashed red rectangle indicates the sky region with NIR data from King \e
(2013), while the southeastern region delimited by the dashed orange
line has only VVV NIR data.
\label{gaia-spatial}}
\end{figure}

\section{Observational data}
\label{obs}

This study covers a square sky region of about 1.5$^{\circ}$ on a side,
($270.2<RA<271.9$; $-25.1<Dec<-23.6$)
including the \ngc\ cluster and the associated Lagoon Nebula in their
entirety, as well as a wide surrounding region. This choice enables us
to search for a possible diffuse population of the cluster, in
addition to that found in its densest regions. It is also useful in
establishing, through the study of the extinction spatial pattern, the
shape and boundaries of the cloud which originated the cluster, and its
environment. This star-formation region is a known ``blister'',
excavated by winds and radiation from its massive stars (e.g., Damiani
\e 2017a). This type of geometry is also suggested by the extinction
pattern, which is seen as a lower star density in correspondence to
thicker dust, for instance in the Gaia (DR2) source catalog whose
spatial distribution is shown in Fig.~\ref{gaia-spatial}. In the same
figure we show the Chandra FOVs, which only covered the central regions:
cluster core (left-hand FOV) and Hourglass (right-hand FOV).
Near the center of the Hourglass region, the Gaia source density is
lowest since the dust has the largest column density and obscures
background stars; the Hourglass nebula is also a bright diffuse nebula,
which causes the optical limiting magnitude to shift towards brighter
values. In general, however, the star-density pattern does not follow the
shape of the bright nebulosity, so that we conclude that the latter has
a minor effect on apparent star density compared to dust extinction.

\begin{figure}
\resizebox{\hsize}{!}{
\includegraphics[angle=0]{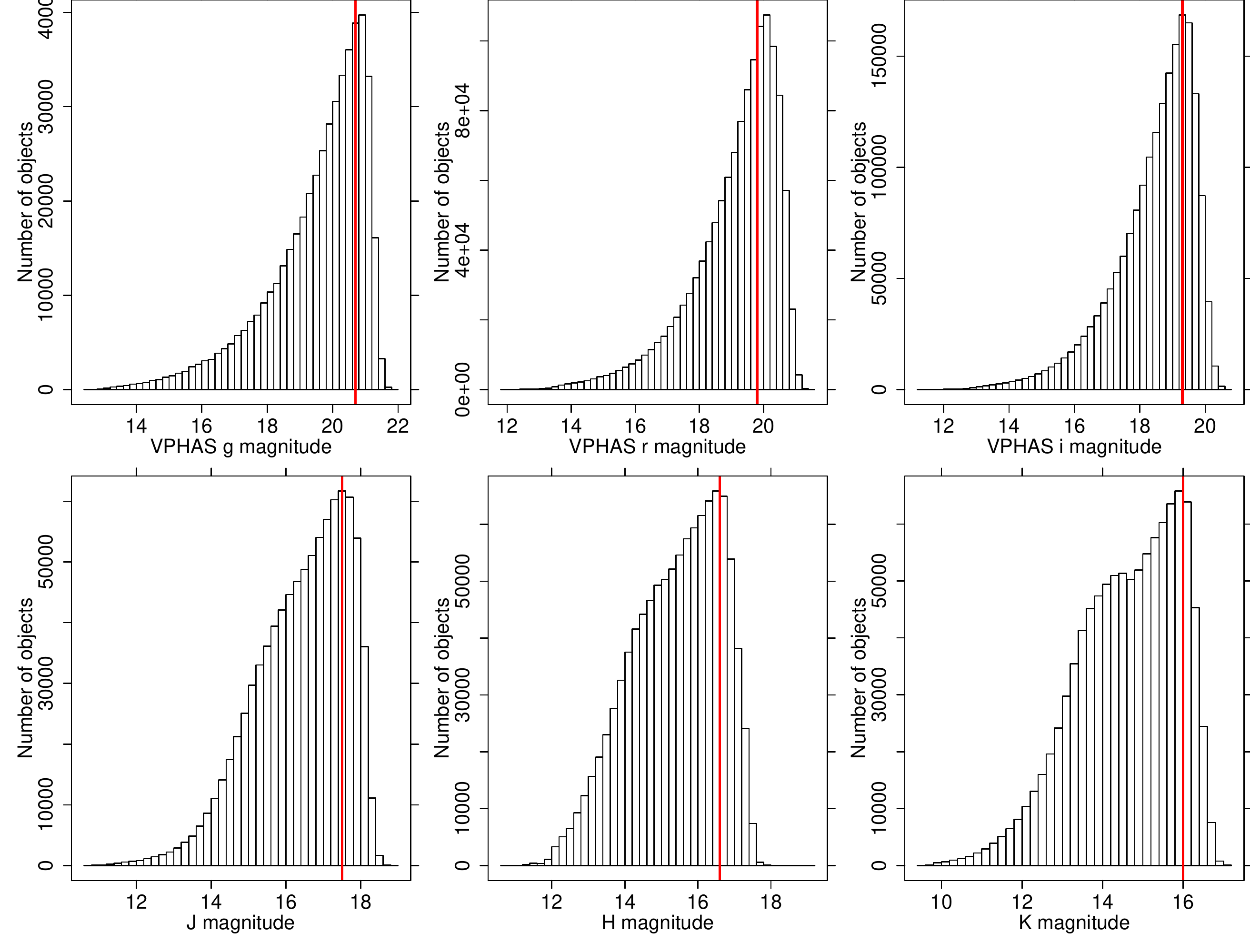}}
\caption{
Magnitude histograms for VPHAS+ bands $g,r,i$ (top panels) and NIR bands
$J,H,K$ (bottom panels). The vertical red lines indicate our adopted
completeness limits.
\label{grijhk-hist}}
\end{figure}

\begin{figure*}
\resizebox{\hsize}{!}{
\includegraphics[angle=0]{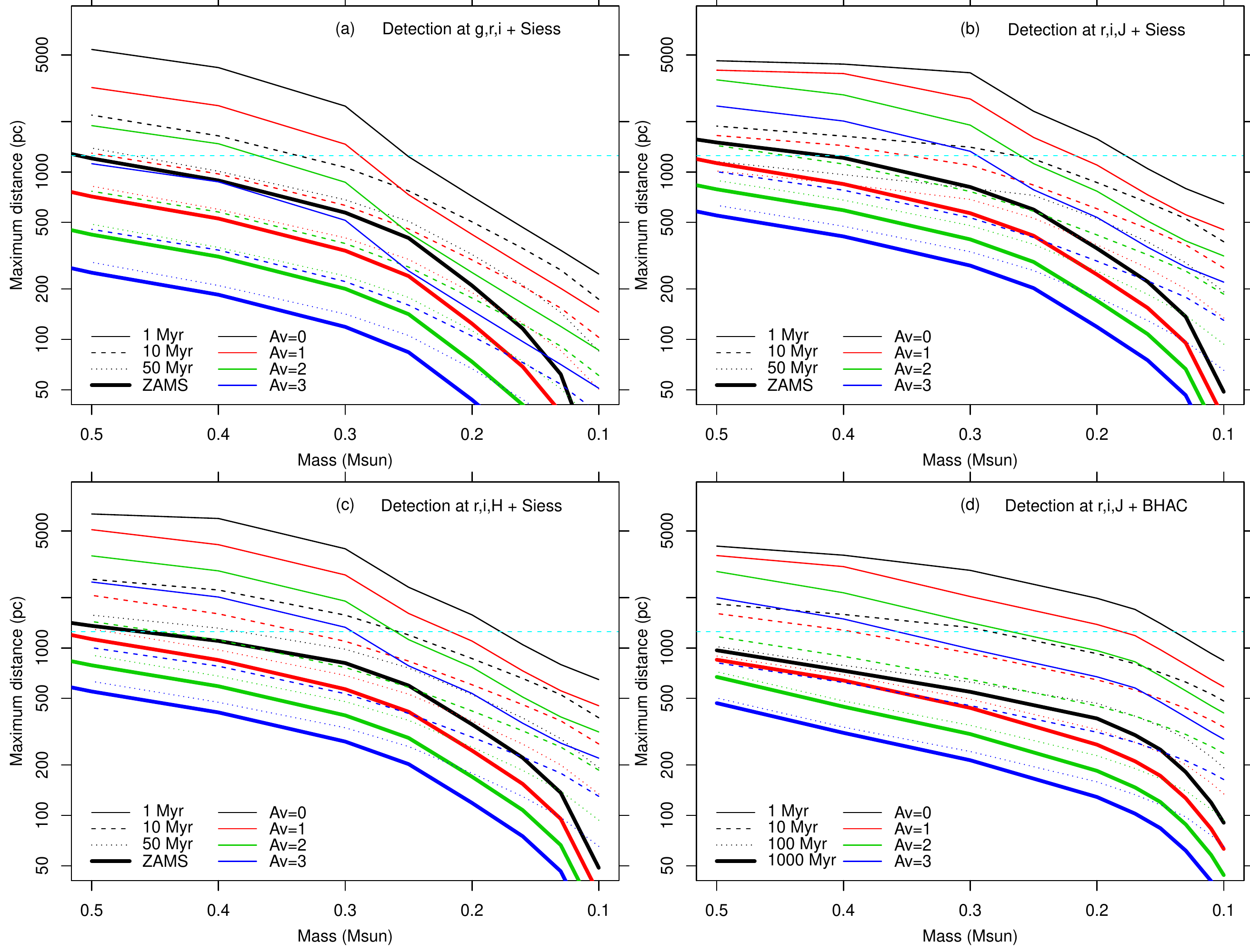}}
\caption{
Mass-Distance-Age (MDA) diagrams:
Maximum distances
for simultaneous detection in three bands,
using completeness limits from Fig.~\ref{grijhk-hist}, and PMS evolutionary
tracks from Siess \e (2000) or BHAC, as indicated.
The cyan horizontal dashed lines indicate the distance of the NGC~6530 cluster.
\label{max-distance}}
\end{figure*}

\subsection{Optical and NIR photometry}
\label{photom}

We use photometric data from the VPHAS+ survey of the Galactic Plane, in
the $u$, $g$, $r$, $i$ and (narrow) \ha\ bands. The VPHAS+ (DR2)
spatial coverage of the studied region is complete, and the number of
VPHAS+ objects found is 2109762. As in previous works
(e.g., Kalari \e 2015, Damiani \e 2016, Paper~I), the $r-$\ha\
index will be used to select accreting PMS stars with strong \ha\ lines,
while the $u-g$ index is useful to find PMS stars with UV-excesses arising
from the accretion spot. 
Magnitude histograms and completeness limits for the $g$, $r$, and $i$ bands
($g=20.7$, $r=19.8$, $i=19.3$)
are shown in Fig.~\ref{grijhk-hist}. The bright limit for VPHAS+ data
is $i \sim 12$, corresponding to cluster stars with masses $\sim 3-4
M_{\odot}$. The few members brighter than this limit, and missed using this
photometric set, are however recovered using X-ray or Gaia data (see below).
VPHAS+ sources are only considered if they have a {\it clean}
flag set for at least one band; magnitude values without a {\it clean}
flag set for the given band are not used.

The assembly of our NIR catalog (in the $JHK$ bands) was not as simple as for
VPHAS+: the region was not completely covered by the UKIDSS Galactic Plane
Survey, while 2MASS is too shallow for the purpose of detecting M stars at the
\ngc\ distance. The central parts of the studied region
(see Fig.~\ref{gaia-spatial}), however, were
observed by King \e (2013) using WFCAM at UKIRT, that is the same
instrumentation used for UKIDSS surveys; therefore we used the catalog
from King \e (2013). We did not use magnitude values with listed errors larger
than 0.3~mag, nor with flags equal to 'S', 'F', 'N', 'E', or 'M'.
The resulting NIR catalog contains 614787 objects, uniformly covering the
region between $270.4977<RA<271.4697$ and $-24.63271<Dec<-23.7505$.
In the region covered by both the King \e (2013) and the UKIDSS catalogs
we matched both catalogs to compare their properties. Using
the 157811 sources common to both catalogs, we found no significant
difference in the respective magnitude values. Magnitude errors in King
\e (2013) were however significantly smaller than in UKIDSS. Therefore,
for the spatial region covered by both, we used the King \e (2013)
catalog. Outside that region we used instead the UKIDSS catalog (303481
entries), keeping only (star-like) sources with flag {\it mergedClass=-1}
or {\it -2}, and magnitudes with flags {\it jppErrBits<255}, {\it
hppErrBits<255}, and {\it k\_1ppErrBits<255}, for the $J$, $H$ and $K$
bands respectively.

\begin{figure*}
\centering
\includegraphics[width=16.5cm]{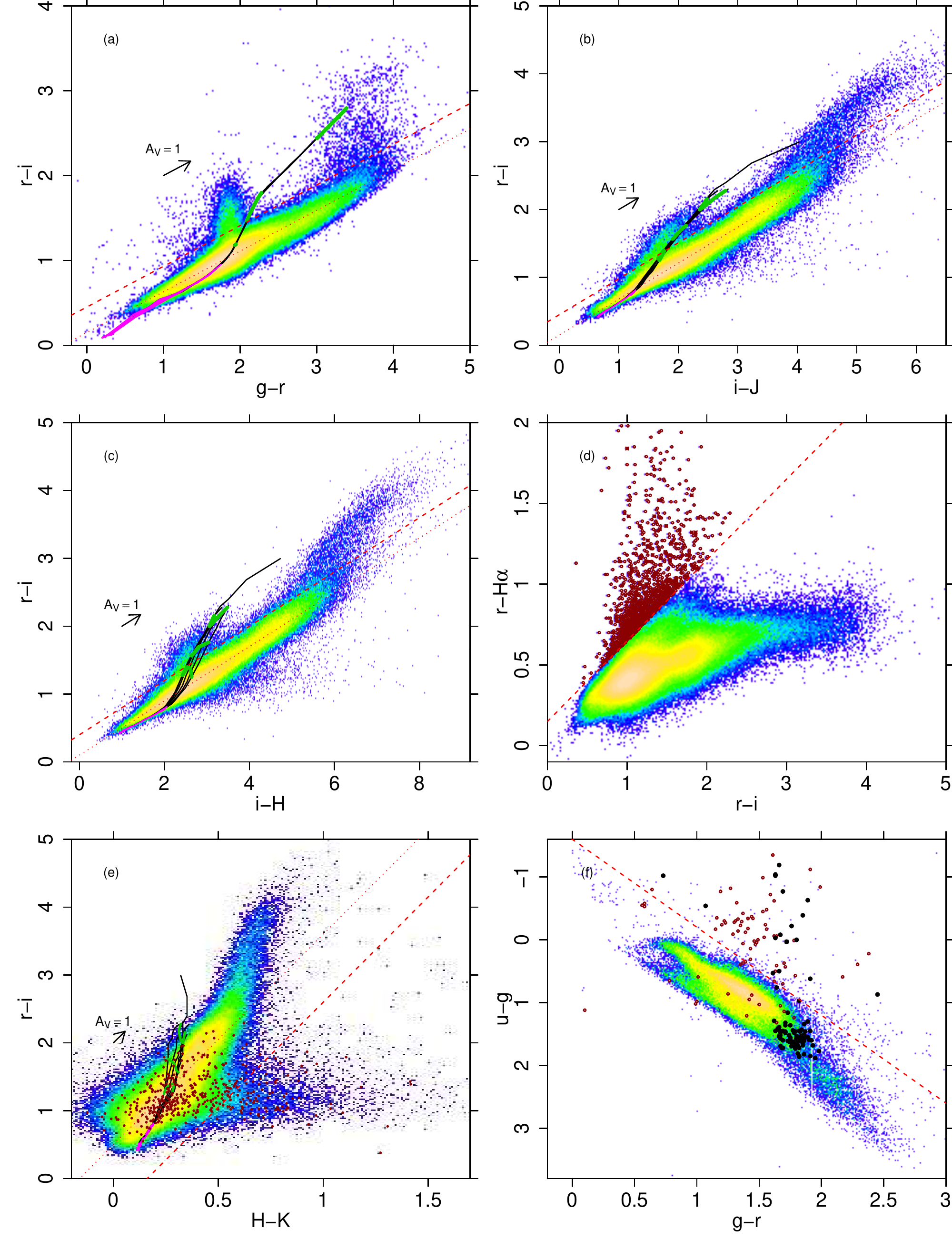}  
\caption{
Color-color diagrams used for selection of young stars.
Each diagram is a two-dimensional histogram, with colors indicating the
density of datapoints.
Only data with errors less than 0.1~mag on each pair of colors are shown.
The dotted lines, parallel to the reddening vector (except for panel
$(d)$ where no reddening vector is defined), describe the bulk of datapoints.
Panels $(a)$,
$(b)$ and $(c)$: M stars are found above the dashed lines, parallel to
the reddening vectors. Panel $(d)$:
strong \ha-emission stars are found above the dashed line (dark-red points).
Panel $(e)$:
IR-excess stars are found below the dashed line
(dark-red points are the same as in panel $(d)$).
Panel $(f)$:
$(g-r,u-g)$ color-color diagram for selection of UV-excess stars
(above the dashed line).
Black dots indicate M stars, while dark-red points are as in panel $(d)$.
In panels $(b,c,e)$,
BHAC isochrones (evolutionary tracks) are shown with black (green) lines for
ages 1, 10, 50 and 10,000~Myr (masses of 0.1, 0.3, and 0.5 $M_{\odot}$),
reddened as appropriate for NGC~6530 {($A_V = 1.085$)}.
The location of stars with mass $M>1 M_{\odot}$ (all ages) is indicated
with a magenta curve. In panel $(a)$ only, Siess \e
(2000) isochrones and tracks are used, the BHAC set being unavailable
for the $g$ band.
\label{definitions}}
\end{figure*}

Still, the southeastern part of the whole studied region
(Fig.~\ref{gaia-spatial}) remain
uncovered by the combined King-UKIDSS NIR catalog. This region was
(almost) completely covered by the VVV survey in the same bands, so we
adopted this latter catalog for the remaining spatial region. We found a
systematic offset between the UKIDSS and VVV $H$ and $K$ magnitude
values { (see Appendix~\ref{append1})},
so before merging the catalogs we subtracted 0.05~mag from the
VVV $H$ magnitudes, and 0.02~mag from the VVV $K$ magnitudes. On
average, magnitude errors are much larger in the VVV catalog than in
UKIDSS, for the common objects in the studied field.
We filtered rather heavily the VVV catalog, as described in Paper~I;
we kept 192541 sources in the region of interest.  Magnitude
histograms and completeness limits ($J=17.5$, $H=16.6$, $K=16.0$) for
the King-UKIDSS NIR catalog are also shown in Fig.~\ref{grijhk-hist}.
This King-UKIDSS-VVV catalog contains 1110809 entries.

For bright NIR sources all the above catalogs are saturated. Therefore, 
we used the 2MASS catalog (275607 sources in the whole region), filtered
as in Paper~I, and matched it within 0.2~arcsec with the
King-UKIDSS-VVV catalog. The number of unique NIR sources in the
obtained King-UKIDSS-VVV-2MASS catalog (NIR catalog for brevity) is 1265426.
Before merging, we added small zero-point offsets
to the King-UKIDSS-VVV magnitudes,
for better matching with 2MASS magnitudes, of
+0.0675~mag in $J$, -0.043~mag in $H$, and -0.0033~mag in $K$.

\begin{figure*}
\resizebox{\hsize}{!}{
\includegraphics[]{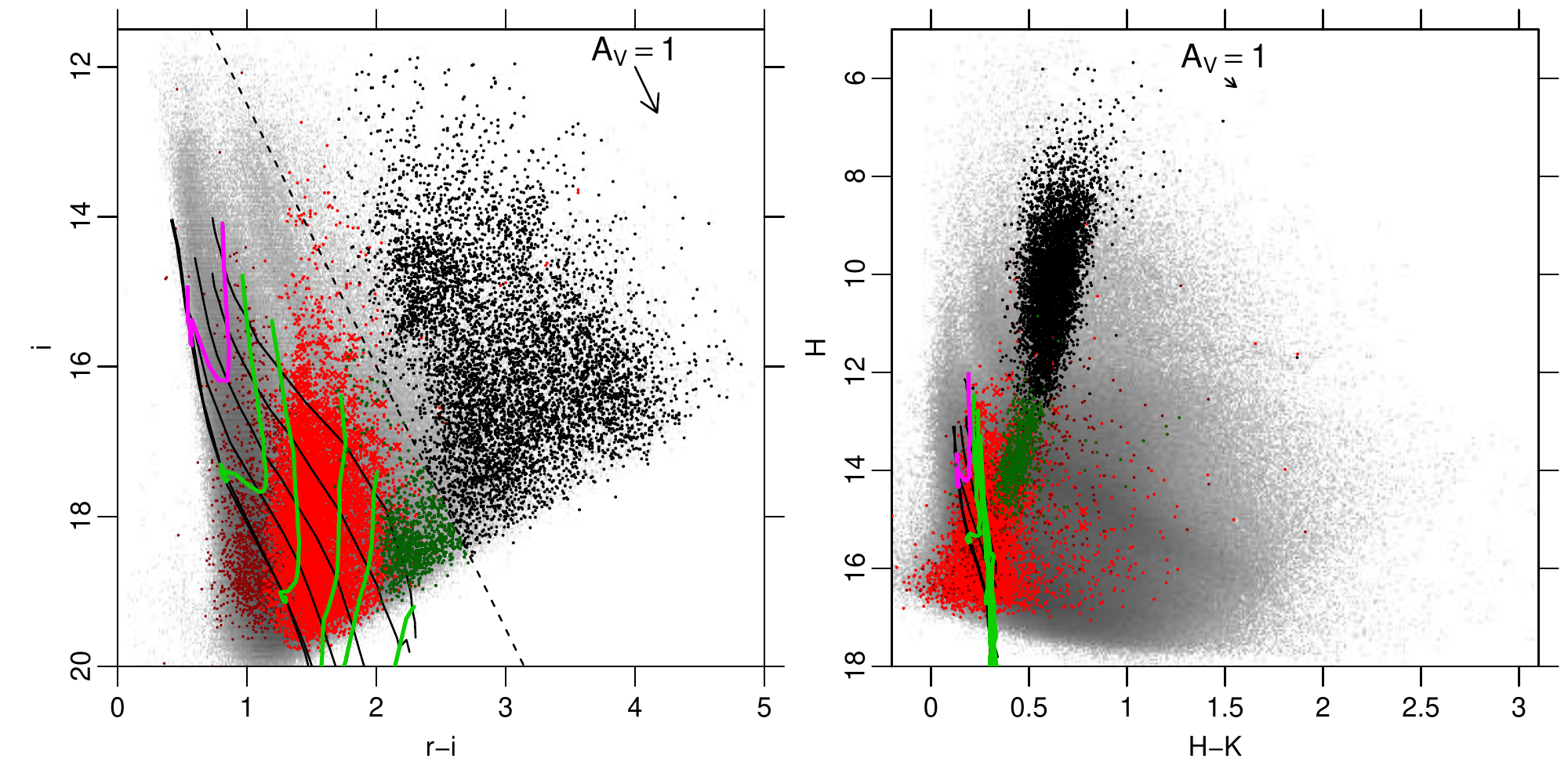}}
\caption{
Color-magnitude diagrams for all stars in the studied region (left:
$(i,r-i)$; right: $(H,H-K)$).
Only data with errors less than 0.1~mag on colors and magnitudes are shown.
Tracks and isochrones { (ages of 1, 3, 10, 30, 100, 10000~Myr)}
as in Fig.~\ref{definitions},
assuming a distance of 1250~pc and extinction $A_V=1.085$.
The magenta isochrones correspond to 1~$M_{\odot}$ stars.
Dark-red points indicate \ha-emission stars, 
light-red dots low-reddening M stars, and black dots high-reddening M
stars (giants).
Dark-green dots are high-reddening M stars which by virtue of their
position in the CMD are included in the low-reddening sample.
\label{cmd}}
\end{figure*}

By matching the NIR and the optical catalogs within 0.2~arcsec we
obtained our final optical-NIR catalog, with 2909548 unique entries.
Of them, 470178 sources have both optical and NIR photometry.

\subsection{Astrometric data}
\label{gaia}

Gaia DR2 data include position, proper motion, parallax, and magnitudes
in three bands ($G$, $BP$, $RP$) for 588823 sources in the studied
region, distributed in space as shown in Fig.~\ref{gaia-spatial}.
They were matched as above with the optical-NIR catalog, yielding
2950266 unique optical-NIR-Gaia sources. The number of sources common to
Gaia and the optical-NIR catalog is 548105. We applied no filtering to
the Gaia data at this stage.

\subsection{X-ray data}
\label{xdata}

We use also X-ray data from the Chandra X-ray Observatory. As mentioned
above, Chandra observed two FOVs in { the Lagoon Nebula}
(Fig.~\ref{gaia-spatial}).
Results from the ``cluster core'' field (ObsID 977, 60~ks exposure
time) were published by Damiani \e (2004), while the combined analysis
of this same dataset and the much deeper Hourglass field (ObsIDs 3754,
4397, and 4444, sharing the same FOV and totalling 180~ks exposure) was
performed as part of the MYStIX project, among several other massive
star-formation regions (Feigelson \e 2013, Kuhn \e 2013). In general,
choices regarding source detection thresholds were more conservative in
Damiani \e (2004) than in the MYStIX papers; in the latter, the existence
of very faint X-ray detections almost requires an independent confirmation
from identification with another catalog (typically in the NIR). The fact
that the Hourglass Chandra field is much deeper than the cluster-core
field also mean that X-ray sources in the former are overrepresented,
and may give a biased picture of the actual stellar density distribution across
sub-regions { in the Lagoon Nebula} because of the spatially
non-uniform depth.
We adopt here the MYStIX catalog from Kuhn \e (2013),
including 2427 X-ray sources, but in practice restrict our
considerations exclusively to the subset of 1756 sources with an optical,
NIR, or Gaia counterpart, for the reasons explained above.
{ The positional match between X-ray sources and the other catalogs was
made within $2.5\times \sigma_\mathrm{Xpos}$ summed in quadrature with
0.07~arcsec, where $\sigma_\mathrm{Xpos}$
is the X-ray position error, which varies from source to source because
of counting statistics and PSF variations across the Chandra FOV, and
0.07~arcsec was adopted as a safety minimum match radius.
}

\section{Member selection}
\label{members}

\subsection{M-type stars}
\label{mstars}

We have shown in Paper~I the effectiveness of the selection of
PMS M-type stars based on multi-band photometry. M stars are selected
from three color-color diagrams (the $(r-i,g-r)$, $(r-i,r-J)$, and
$(r-i,r-H)$ diagrams), where they lie along a locus which removes the
$T_{eff}-A_V$ degeneracy. M stars remain therefore well recognizable,
unlike stars of other spectral types, at all values of extinction.
Moreover, since PMS M stars are much brighter than main-sequence M
stars, they are observable up to much larger distances for given
magnitude limits (as appropriate for a particular photometric dataset),
and this makes a { cluster rich in PMS M stars} recognizable against a more
diffuse field-star population of nearby M stars. For details, see Paper~I.
Quantitative predictions on detection limits for the dataset
being studied here were made in the same way as in Paper~I, on
the basis of theoretical isochrones from Siess \e (2000) and Baraffe \e
(2015; BHAC), the completeness limits for each photometric band, and
knowledge of the reddening law
{ (see Appendix~\ref{append2}).}
For each value of age and extinction
$A_V$ there is a minimum mass for a star to be detected in the
photometric catalog, for a given distance of the cluster.
Alternatively, for a given mass there is a maximum distance for
detection for given age and extinction.
These relations are graphically expressed by the Mass-Distance-Age (MDA)
diagrams, already used in Paper~I, and reproduced for the case of \ngc\
in Fig.~\ref{max-distance}. These diagrams differ subtly from those in
Paper~I, since the magnitude limits in the $J$ and $H$ bands are not the
same, and since the reddening law is found here to differ slightly from
that in Sco~OB1 (see below). Moreover, we are here advantaged by
the smaller distance of \ngc\ compared to Sco~OB1.
The two diagrams on the right-hand side of Fig.~\ref{max-distance}
differ only in the theoretical tracks used, and give an idea of the
model-related uncertainties in the method; as is seen, these affect the
details but not the basic pattern of predicted detections versus mass,
distance, extinction, and age.
{ As an example, the minimum detectable mass in $r,i,J$ at the
NGC~6530 distance, 1~Myr age, and $A_V=1$ ($A_V=2$) is 0.22 (0.26) $M_{\odot}$
according to Siess models, while is 0.18 (0.27) $M_{\odot}$ according to BHAC.
At 10~Myr and $A_V=0$, Siess and BHAC predict minimum masses of
respectively 0.26 and 0.285 $M_{\odot}$.
}

The color-color diagrams used to select M stars from our optical-NIR
catalog in the { Lagoon Nebula} region are shown in the first three panels of
Fig.~\ref{definitions}. The large number of field stars (mostly
background) contained in the catalog, spanning a very wide range of
extinction, permits to define accurately the average reddening law
(over a large distance range) in this sky direction. This is described by
the dotted lines, fitted to the bulk of datapoints. The dashed lines
separate M stars (above them) from all other spectral types.
{
While their slopes are fixed by the reddening vectors, intercepts were
empirically determined as a compromise between maximum inclusion of
M-type stars and minimum inclusion of reddened hotter stars. In all
cases, the maximum mass they select is comprised between 0.3 and 0.5
$M_{\odot}$ (thick green lines in the figure), the exact values being
slightly age-dependent.
}
As also
predicted by the model isochrones shown in the figure, the distribution
of M stars does not run parallel to the reddening vector in each of the
three $(r-i,g-r)$, $(r-i,r-J)$, and $(r-i,r-H)$ diagrams. This makes
also possible to determine rather accurate extinction values for
individual M stars, by back-projecting them onto the model tracks at
zero extinction (except for the $(r-i,r-H)$ diagram, the model locus is
not age-dependent). The general pattern in all three panels is very
similar to that found in Sco~OB1 (Paper~I), with a low-extinction group
of M stars well separated from another group of M stars at much higher
extinction ($A_V \sim 8-10$~mag), which was argued in Paper~I to be
constituted by distant M giants. Below we will present additional evidence
(other than that provided in Paper~I) that this is indeed the case. The
\ngc\ M-type members are therefore expected to belong to the
low-reddening M star groups (the MDA diagrams can also be used to
confirm that we would be unable to detect any 1-Myr-old M stars in \ngc\
with $A_V>5-6$, using the present catalog).

As in Paper~I, we define three groups of M stars (i.e., above the dashed
line in Fig.~\ref{definitions}$a$), named after the
respective diagrams, as:
\begin{equation}
M_{gri}:\; \; (r-i) > 0.45+0.48\; (g-r)
\label{mgri}
,\end{equation}
and also $1.5<(g-r)<2.5$ to discriminate from high-reddening M giants.
\begin{equation}
M_{riJ}:\; \; (r-i) > 0.45+0.53\; (i-J)
\label{mrij}
,\end{equation}
and $1.2<(i-J)<3.0$.
\begin{equation}
M_{riH}:\; \; (r-i) > 0.4+0.4\; (i-H)
\label{mrih}
,\end{equation}
and $2<(i-H)<4$.
High-reddening M stars (giants) are defined using the same
equations \ref{mgri}~to~\ref{mrih}, but above the limits $(g-r)>2.5$
($gM_{gri}$ stars), $(i-J)>3$ ($gM_{riJ}$), and $(i-H)>4$ ($gM_{riH}$),
respectively. The slopes used in Eq.~\ref{mgri}-\ref{mrih} are the same
as those of the reddening vectors.

The numbers of stars found in each group are 3298 $M_{gri}$, 3547
$M_{riJ}$, and 2970 $M_{riH}$ stars; the latter two are largely
overlapping, with 2631 stars in common, while there are only 696 stars
in common between $M_{gri}$ and the combined $M_{riJ}$ plus $M_{riH}$
group. The total number of low-reddening M stars is 6488.
For the high-reddening giants, the corresponding numbers are 693 $gM_{gri}$,
2987 $gM_{riJ}$, and 2919 $gM_{riH}$ stars, for a total number of 3620 stars;
the overlap beween $gM_{riJ}$ and $gM_{riH}$ samples is nearly complete
(2804 stars), while that between these two and the $gM_{gri}$ stars
includes only 175 stars (the large extinction towards distant giants
preventing $g$ band detection for most of them).

\subsection{\ha\ emission, NIR and UV excesses}
\label{classic}

The $r-$\ha\ index from the VPHAS+ catalog was used to select CTTS from
their strong \ha\ emission,
using the diagram in Fig.~\ref{definitions}$d$, in the same way as it
was done in Paper~I. We find in this way 1254 emission-line stars
over the whole studied region, of which 150 also belonging to the
low-reddening M-star sample. Using { a similar, but more
conservative} selection method, Kalari \e (2015)
found 235 \ha-emission stars inside a $50^{\prime} \times 30^{\prime}$
field in the Lagoon nebula region, { smaller than the region
considered here. Compared to Kalari \e (2015), our selection is more
inclusive of stars near the CTTS threshold.}

Fig.~\ref{definitions}$e$ shows an optical-NIR color-color diagram,
useful to select Class~II PMS stars with disk-related NIR
excesses. As in Paper~I (and already argued in Damiani \e 2006a, 2017b, or
Guarcello \e 2007), this mixed optical-NIR diagram selects more objects
than the $(J-H,H-K)$ diagram (not shown here), being
sensitive even to weaker excesses in spite of requiring data at more
bands. We find 2173 NIR-excess stars selected from
Fig.~\ref{definitions}$e$, of which only 73 in common
with the low-reddening M-star sample, and 79 with the \ha-emission sample.
{
The total number of optical-NIR-excess objects is much larger than found in the
analogous study by Damiani \e (2006a - 333 sources), both because the NIR
data used here are much deeper that the 2MASS data of the former study,
and because of the much larger field studied here, despite that Damiani \e
(2006a) used several NIR-NIR and optical-NIR diagrams for the highest
completeness while we here use only the $(r-i,H-K)$ diagram.
}

Finally, Fig.~\ref{definitions}$f$ shows the $(u-g,g-r)$ diagram, useful
to select stars with an UV excess, which in the case of PMS stars is
commonly attributed to the accretion spot where disk material hits the
star surface. We find here 436 UV-excess stars, of which 17 among the M
stars (the overall number of M stars with $u$ band data is anyway very
small - black dots in Fig.~\ref{definitions}$f$,
{ 61 stars}), and 75 among the \ha-emission stars.

\subsection{Color-magnitude diagrams}
\label{cmdiag}

We show in Fig.~\ref{cmd} an optical and a NIR color-magnitude diagram
(CMD) for all stars selected using the above methods as low-reddening M
stars, high-reddening M stars, and \ha-emission stars, respectively.
Again, the overall distribution of datapoints is similar to that found
in Sco~OB1 (Paper~I): neither the M stars nor the emission-line stars
form sequences in the CMD, indicating that most of them are field
objects. In particular, the bulk of \ha-emission stars falls below the
ZAMS at the cluster distance (and nominal extinction), suggesting they
lie farther away. High-reddening M stars are on average brighter than
low-reddening M stars: this is particularly evident in the $(H,H-K)$
diagram in the right panel of Fig.~\ref{cmd} since these bands are less
affected by extinction: this again supports the idea that they are
giants, much brighter than M dwarfs despite their larger reddening and
distance. The $(i,r-i)$ diagram shows a region with a reduced
density of M stars, between the regions occupied by dwarfs and giants
(indicated with a dashed line). A number of stars in the high-reddening sample
(dark-green datapoints) falls below the dashed line, that is closer in the
CMD to the dwarfs than to the giants. We have therefore re-classified
this small group of stars, including them among the low-reddening
M-star sample.

\begin{figure}
\resizebox{\hsize}{!}{
\includegraphics[angle=90]{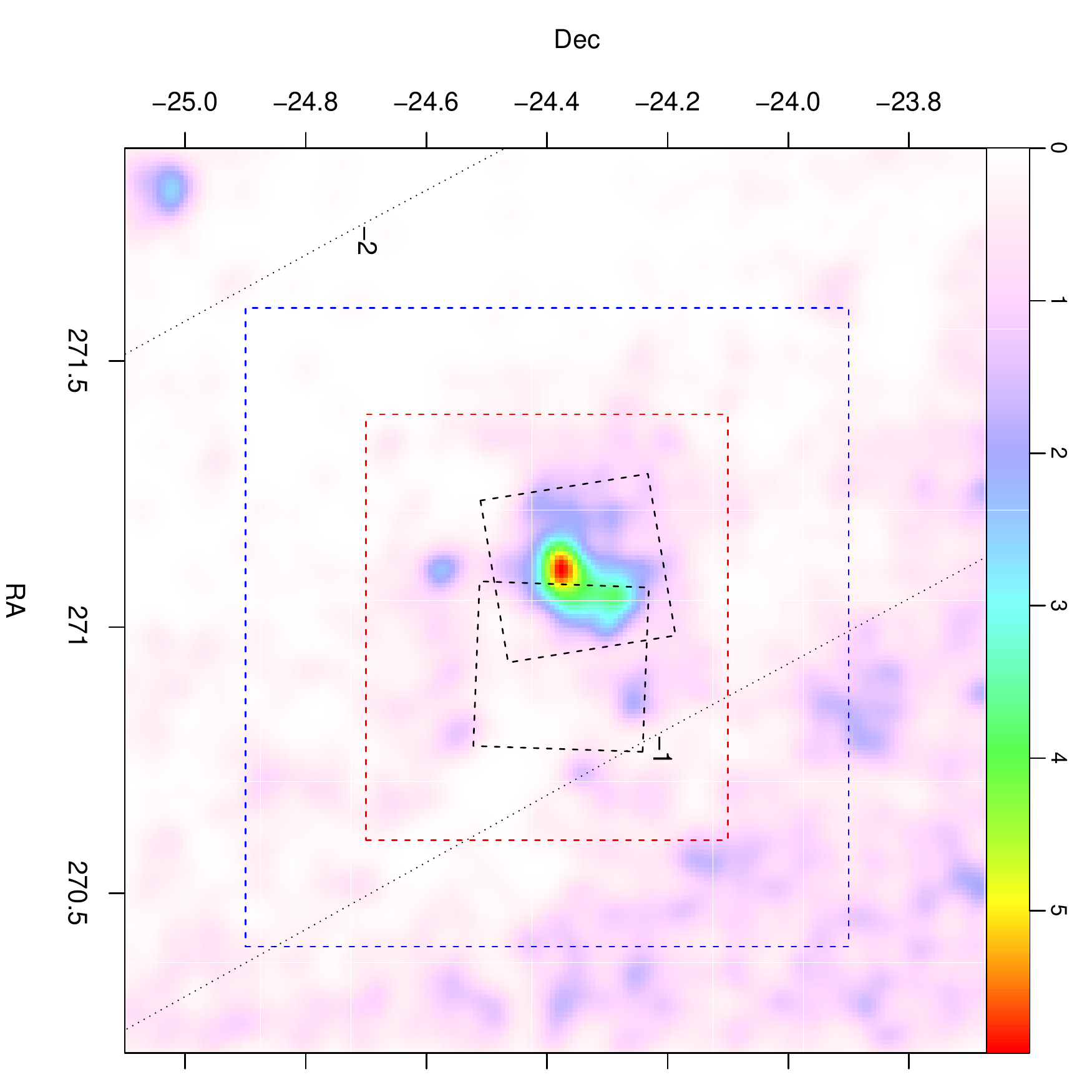}}
\caption{
Smoothed density map of all stars with either \ha\ emission, NIR or UV
excesses. The smaller black squares are the Chandra FOVs as in
Fig.~\ref{gaia-spatial}. The red dashed rectangle is a provisional
cluster region, while the outer blue rectangle indicates the
surrounding reference field-star region.
The oblique dotted lines indicate constant Galactic latitude $b$, as
labeled. The enhancement in the lower-left corner corresponds to the
globular cluster NGC~6544.
Labels in the top-axis colorbar indicate density in units of sources per
square arcmin.
\label{smoothed-halpha}}
\end{figure}

\subsection{Proper motion and parallax}
\label{plx}

We also considered the data on proper motion
$(\mu_{\alpha},\mu_{\delta})$ and parallax $(\pi)$ provided in
the Gaia DR2 catalog (Luri \e 2018), as a further membership criterion.
The \ngc\ cluster is not immediately visible in the proper-motion plane
using the Gaia data, its motion being not dramatically different from
that of the much more numerous field stars in its surroundings.
The same holds for the parallax-proper motion plane (either along
$RA$ or $Dec$). We have therefore used a differential approach, by
defining an initial ``cluster'' region, and a comparison ``reference''
region where we have no indications of the possible presence of cluster
members. The cluster region was chosen from the stars fulfilling any of
the \ha, NIR, and UV selection criteria (named collectively the
candidate CTTS stars). The (smoothed) spatial density of CTTS is shown
in Fig.~\ref{smoothed-halpha}. The enhancement near the center,
corresponding to \ngc\ is evident. A density gradient from SE towards NW
(i.e., towards the Galactic Plane) is also clearly seen, and \ha\ or
NIR-excess stars near the Plane are neither clustered nor related to
\ngc. The enhancement in the SE corner corresponds to the globular
cluster NGC~6544. From the figure, we set the initial cluster region,
enclosing nearly all of its possible candidate members, as
that inside the red rectangle, and the reference region as that between
the red and the blue rectangles.

\begin{figure*}
\resizebox{\hsize}{!}{
\includegraphics[]{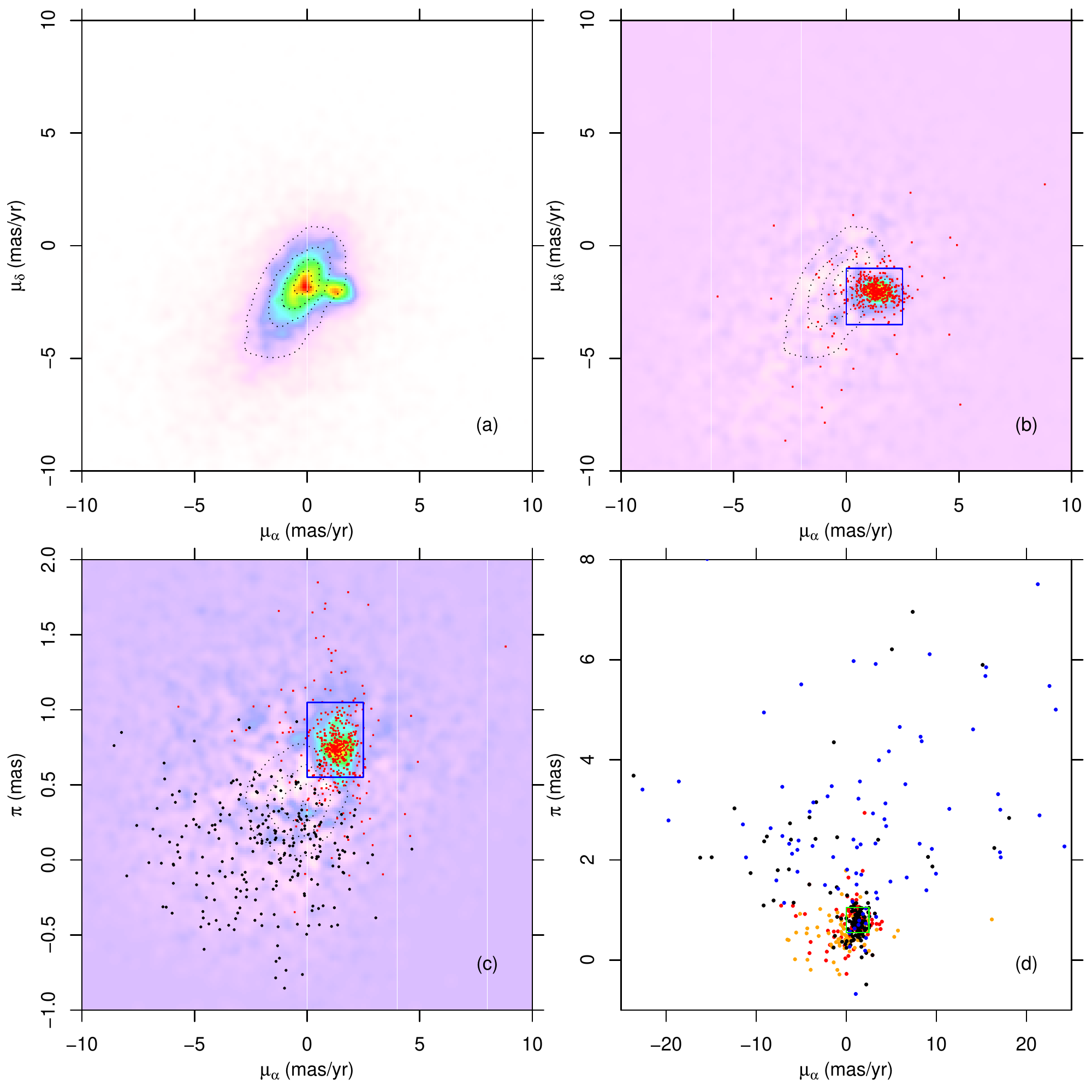}}
\caption{
$(a)$: Distribution in proper-motion plane of all Gaia sources falling
in the ``cluster region'' defined in Fig.~\ref{smoothed-halpha}, and
having errors on $\mu_{\alpha}$ and $\mu_{\delta}$ less than 0.5~mas/yr.
The dotted contours describe instead the proper-motion distribution of
Gaia sources in the ``reference'' region defined in
Fig.~\ref{smoothed-halpha}, excluding those in the cluster region.
$(b)$: Map of density difference between Gaia sources in cluster and
reference regions, scaled by respective total sources numbers, in
proper-motion plane. Red dots are \ngc\ X-ray sources. The blue
rectangle indicate the cluster boundaries in the proper-motion plane.
Contours are the same as in panel~$a$.
$(c)$: Map of density difference between Gaia sources in cluster and
reference regions, scaled by respective total sources numbers, in
$\mu_{\alpha}$-parallax ($\pi$) plane. The maximum parallax error for
inclusion is 0.3~mas.
Red dots are \ngc\ X-ray sources.  Black dots are candidate M giants.
The blue rectangle indicate the cluster boundaries in the
$\mu_{\alpha}$-parallax plane.
Contours describe the distribution of sources in the reference region.
$(d)$: A wider-scale view of the $\mu_{\alpha}-\pi$ plane of panel~$c$.
The green rectangle is the same as the blue rectangle in panel~$c$.
Black (blue) dots indicate low-reddening M stars in the cluster
(reference) region, while red (orange) dots indicate CTTSs in the
cluster (reference) region.
\label{gaia-all}}
\end{figure*}

The density of datapoints in the proper-motion plane, including only
stars in the cluster region, is shown in Fig.~\ref{gaia-all}$a$:
superimposed to the distribution of the field stars (the majority of
datapoints, in a nearly-elliptical distribution centered on $\mu_{\alpha}
\sim -0.253$~mas and $\mu_{\delta} \sim -1.795$~mas/yr)
we see clearly a narrower
density peak in the range $0.0 < \mu_{\alpha} < 2.5$~mas and
$-3.5 < \mu_{\delta} < -1.0$~mas/yr. The density distribution
in the reference field is shown in the same Fig.~\ref{gaia-all}$a$ with
the dashed contours, which only indicate the presence of the field-star
component, as expected. Only sources with errors on both $\mu_{\alpha}$
and $\mu_{\delta}$ less than 0.5~mas are shown.

\begin{figure}
\resizebox{\hsize}{!}{
\includegraphics[]{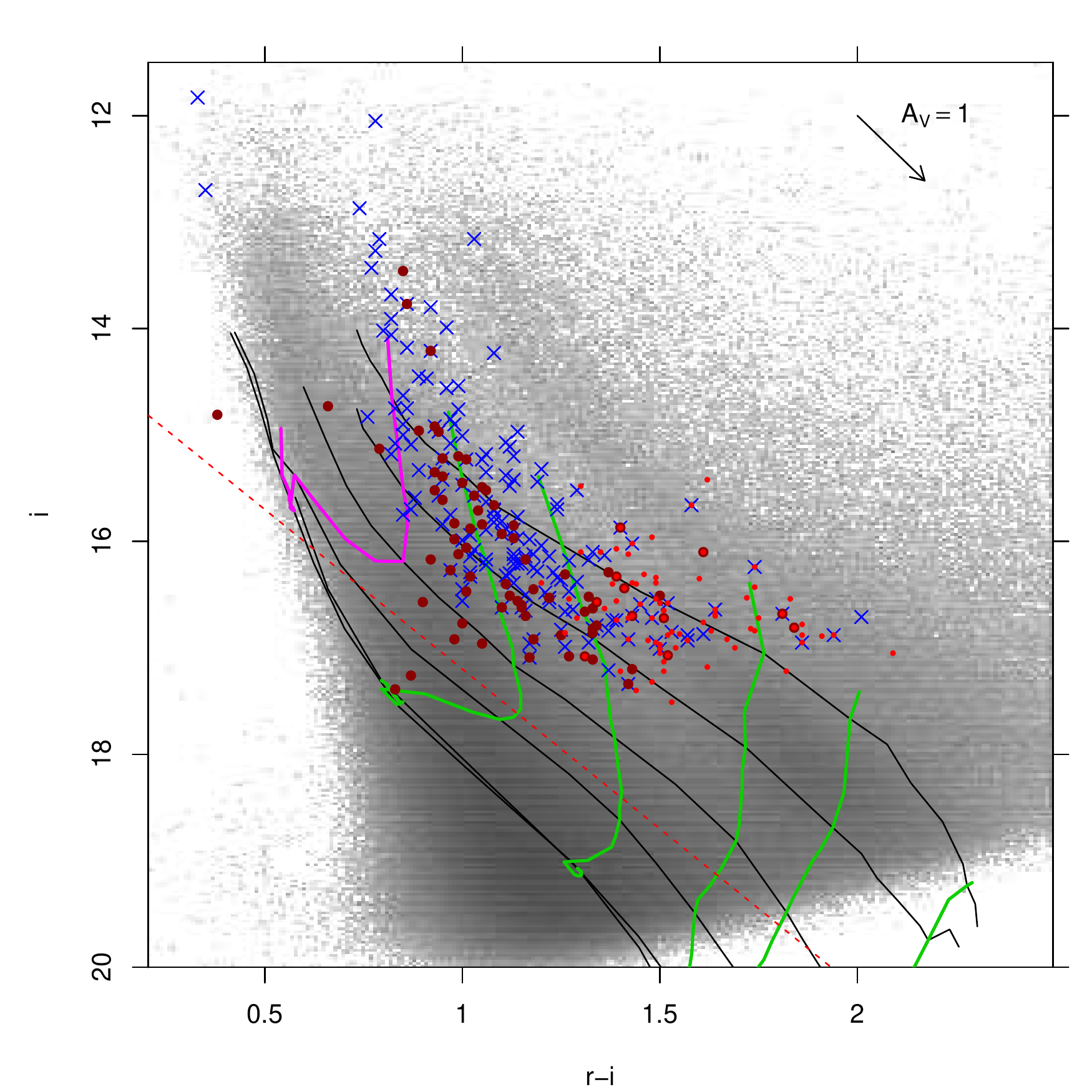}}
\caption{
The $(i,r-i)$ color-magnitude diagram of M stars (red dots), X-ray
sources (blue crosses), and CTTSs (dark-red points) which are also
selected using Gaia proper motions and parallaxes.
Background gray-scale image: same CMD as in Fig.~\ref{cmd}$a$.
Evolutionary tracks, isochrones and reddening vector as in the same figure.
The red dashed line (roughly corresponding to a 25-Myr isochrone)
delimits the CMD region occupied by X-ray and M-type members.
\label{cmd-gaia-sel}}
\end{figure}

In order to better isolate the cluster in the proper-motion plane we
normalize both the cluster and reference distributions to unity (to
account for the difference in the respective number of stars), and plot
their difference in Fig.~\ref{gaia-all}$b$. This fluctuates around zero,
except for the cluster stars which form a well-defined density peak
(outlined with a blue rectangle). It should be clear that the
normalization procedure is only an approximation, both since it does not
take into account the expected larger density of stars in the cluster
region because of the cluster contribution, and since the non-uniform
spatial obscuration does not guarantee that the properties of the field
stars in the two regions compensate exactly. The map of
Fig.~\ref{gaia-all}$b$, moreover, has by construction zero average, so
that the positive cluster peak must be balanced by a sparser distribution
of negative values (the bluest ones). Nevertheless, this representation
serves its main purpose very well, the cluster peak being evidently
well defined. In the same panel, we show all X-ray detections with a
match in the Gaia catalog: a negligibly small percentage of them lies
outside the blue rectangle (the ``cluster box''), well compatible with
either spurious X-ray-Gaia associations or field X-ray sources. The
bulk of X-ray detections, which are good candidate cluster members,
confirm that the excess density peak in the proper-motion plane does
indeed correspond to \ngc.

\begin{figure*}
\resizebox{\hsize}{!}{
\includegraphics[]{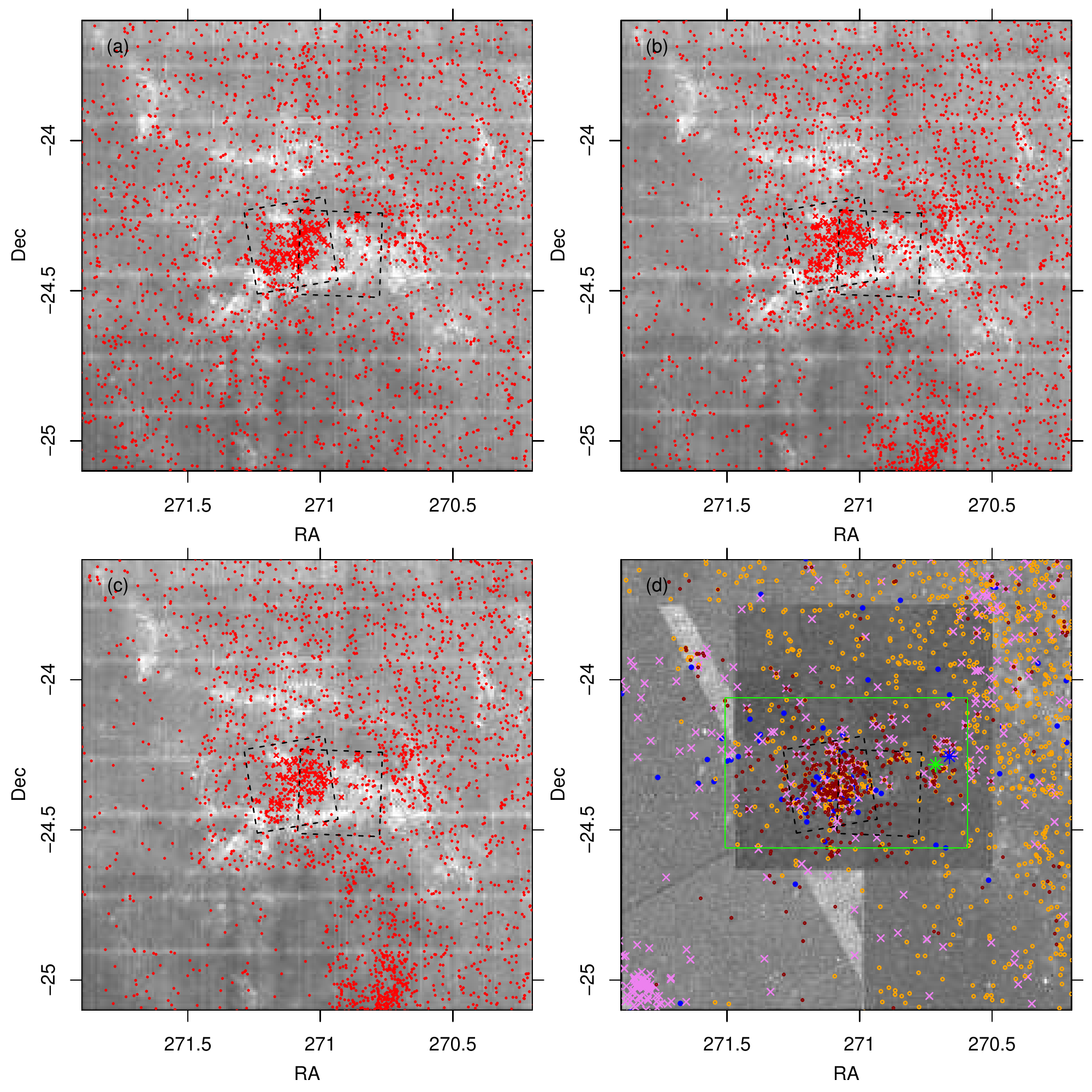}}
\caption{
Spatial distributions of star subsamples in the studied region. The background
image is a two-dimensional density histogram of stars in VPHAS+ DR2
(and sources in our NIR catalog in the bottom right panel).
The dashed squares are the Chandra FOVs.
Top left: $M_{gri}$ stars (red) selected from
Fig.~\ref{definitions}$a$.
Top right: $M_{riJ}$ stars (red) from Fig.~\ref{definitions}$b$.
Bottom left: $M_{riH}$ stars (red) from Fig.~\ref{definitions}$c$.
Bottom right: dark-red points, orange circles, and larger purple crosses
are stars with \ha\ emission, optical-NIR excesses, and UV-excesses,
respectively, as in Fig.~\ref{definitions};
big blue dots are O or B stars from SIMBAD.
The green rectangle is the field studied by Kalari \e (2015).
The green and blue asterisks are the
{ stars 7~Sgr (F2II-III) and HD~164536 (O7.5V),}
respectively.
\label{spatial-Mstars}}
\end{figure*}

The same differential procedure was applied to the $(\mu_{\alpha},\pi)$
plane, with the result shown in Fig.~\ref{gaia-all}$c$: there is a
well-defined peak in the differential density map for $0.55<\pi<1.05$
(blue rectangle), with X-ray detections confirming its coincidence with the
cluster. Only sources with error on $\pi$ less than 0.3~mas/yr are
shown, and proper-motion error less than 0.5~mas as above.
In the same panel, we also show all high-reddening M stars as
black dots: virtually none of them lies inside the cluster ``box'', nor
do they form any cluster in this plane; their parallaxes are on average
much smaller (often consistent with zero\footnote{See Luri \e (2018)
regarding the meaning of negative $\pi$ measurements.}) than for \ngc\ stars,
which confirms that they lie at much larger distances, and we are
correct in our arguments that they must be luminous and distant giants.

Finally, Fig.~\ref{gaia-all}$d$ shows the same density difference as
panel $c$, but uses wider boundaries in both axes, to show not only the
low-reddening M stars in the cluster region (most of them inside the
cluster box, in agreement with our arguments on M-type member selection),
but also the sparse population of field M (dwarf) stars in the reference
region (blue dots), which as expected from our earlier arguments
are found scattered over a large range of $\mu_{\alpha}$ values, and at
definitely larger $\pi$ values (smaller distances) than cluster stars:
the vast majority of M-type dwarfs are too faint to be detected in our
dataset at the \ngc\ distance, as predicted by Fig.~\ref{max-distance}.
Panel $d$ also shows the distribution in the $(\mu_{\alpha},\pi)$ 
plane of the candidate CTTS from the cluster and reference regions:
while those in the cluster region are mostly found inside the cluster
box, those in the reference region are scattered more widely, and often
at smaller $\pi$, indicating that they constitute an unrelated
population of emission-line or NIR-excess objects (so that the CTTS
designation may only be meaningful for the objects inside the cluster region).

Therefore, we consider as Gaia-selected candidate members all sources with
$0.55<\pi<1.05$~mas, $0.0 < \mu_{\alpha} < 2.5$~mas/yr, and
$-3.5 < \mu_{\delta} < -1.0$~mas/yr, and errors on $\pi$ less than
0.3~mas, and errors on $\mu_{\alpha}$ and $\mu_{\delta}$ less than 0.5~mas/yr.

\subsection{Refining photometric selection}
\label{photlim}

By combining several of the previous member selection methods, we are
able to determine a highly reliable subsample of cluster members, from
which we in turn derive other properties of the whole population, and
additional constraints. In particular, \ngc\ members occupy a
well-defined portion of the optical CMD, as is shown in
Fig.~\ref{cmd-gaia-sel} where we only show members selected using Gaia
plus one of the other methods. The complementarity of the methods is
evident from the fact that the respective samples tend to populate
different parts of the CMD, with limited overlap between them.
{
As expected, X-ray detections become rarer towards the lowest masses
because of sensitivity limitations and X-ray luminosity scaling with
mass (e.g., Damiani \e 2006b and references therein), which emphasizes
the importance of a method for selection of the cluster M stars. On the
other hand, CTTSs are scattered across the entire mass range, and
in some cases at apparent ages older than the rest of the cluster
members: this fact was already noticed both in \ngc\ (Damiani \e 2006a)
and in other young clusters (NGC~6611, Guarcello \e 2007; NGC~6231,
Damiani \e 2017b) and is probably unrelated to the stellar ages.
}
We observe that no X-ray or M-type stars are confirmed as members using Gaia
below a certain limit, shown as a red dashed line in the figure, which
approximately corresponds to a 25~Myr isochrone.
Stars below that limit in the CMD are thus very unlikely to belong to
the cluster, unless they show strong spectral peculiarities
(in fact, only two CTTSs fall below it in Fig.~\ref{cmd-gaia-sel}).
We therefore apply this additional constraint to our M-star selection,
which removes 1083 main-sequence M stars at distances close to \ngc.
{
The same constraint was applied to CTTSs, also considering that the
CTTSs falling below the red dashed line in Fig.~\ref{cmd-gaia-sel} show
no spatial clustering. The number of CTTSs reduces therefore to 513
stars.
}

All \ngc\ candidate members selected using the various methods described
are listed in Table~\ref{table-members}.

\section{Results}
\label{results}

\subsection{Distance}
\label{distance}

As can be seen from Fig.~\ref{gaia-all}$c$ the $\pi$ distribution of the
cluster stars is rather wide: this was not unexpected and consistent
with the current precision of parallax measurements in Gaia DR2, also
considering that the majority of cluster members are faint stars and
precision on $\pi$ degrades with magnitude. The actual distribution of
distances expected among cluster members cannot justify a range in $\pi$
from 0.55 to 1.05 ({ nominally,} 950 to 1820~pc).
{ The Gaia data, therefore, do not resolve the cluster along the line
of sight, and by consequence we may only}
derive an
average value for the cluster distance, using a suitable average of
$\pi$ values of its best candidate members. Considering the subsample of
very good candidate members { which are} both X-ray detected and falling
inside the cluster box in both proper motion and parallax (292 stars), their
median parallax is 0.75015~mas.
{ The crucial issue in converting parallaxes to distances is that the
error distribution on the former is symmetric and Gaussian, while that
on distances computes simply as $d=1000/\pi$ is asymmetric, and the
amount of asymmetry depends on relative error on parallax, $\Delta
\pi/\pi$. If $\Delta \pi/\pi$ is sufficiently small (precise $\pi$
measurements), say $\Delta \pi/\pi<0.1$, the asymmetry introduced by
inversion is negligible, and $d=1000/\pi$ is a reliable distance
estimate (see Luri \e 2018, Bailer-Jones \e 2018). As recommended by
these authors, modeling should be done as far as possible directly using
parallaxes, and conversion to distances should be the last operation
performed; parallax errors being Gaussian, error propagation is much
more accurate using parallaxes as well. Therefore, we compute first the mean
parallax of \ngc\ members and its error, and then convert them to
distance and its error.
}
The error-weighted mean parallax for the
{ best candidate members}
is 0.75449~mas, very close to the median, with an error on the
mean (not the standard deviation of datapoints) equal to 0.00412~mas.
{ Therefore, for the cluster as a whole we obtain $\Delta \pi/\pi = 0.0055$,
definitely small enough for direct distance computation using $d=1000/\pi$.}
The corresponding distances are 1333~pc for the median parallax and
$1325^{+7.3}_{-7.2}$~pc for the weighted mean, the distance error being
slightly asymmetric whereas the error on $\pi$ is symmetric (Luri \e 2018).
While very precise, this distance determination is likely not accurate:
Arenou \e (2018) report that Gaia DR2 parallaxes show systematic
residuals (compared to reference values) of $\sim -0.065$~mas, that is
the Gaia distances seem to be overestimated, by $\sim 113$~pc in the
case of NGC~6530.
Were it not for this effect, the Gaia data would locate the
cluster 75~pc more distant than found by Prisinzano \e
(2005 - 1250~pc), but taking into account the above Gaia systematic error the
new Gaia distance is well in agreement with the previous value.

\subsection{Spatial morphology}
\label{spatial}

We discuss here the spatial distribution of candidate \ngc\ members
according to the different membership criteria from the previous
Section. Fig.~\ref{spatial-Mstars}-$a,b,c$ show the distributions of
$M_{gri}$, $M_{riJ}$, $M_{riH}$ stars, respectively.
Fig.~\ref{spatial-Mstars}$d$ shows instead the spatial distributions of
CTTS candidates (i.e., \ha-emission, NIR- and UV-excess stars).
The analysis of the
$M_{gri}$ stars benefits from the uniform catalog coverage across the
entire region studied, and shows clearly a clumpy distribution near the
cluster center (with the largest concentration in the cluster core
region, inside the eastern Chandra FOV), and a uniform distribution in
the diffuse field. Stars in the latter group are field M dwarfs, as also
suggested in Sect.~\ref{plx}, found at smaller distances than the \ngc\
cluster. The effect of mixing different NIR catalogs is clear in panels~$b$,
$c$, and~$d$, with a sharp density drop at their interface, because the
filtering applied to the VVV data made this catalog less complete than
the King-UKIDSS catalog in the central regions. Within these
limitations, the pattern of $M_{riJ}$ and $M_{riH}$ stars in panels~$b$ 
and~$c$ is consistent with that derived from panel~$a$, except for a
conspicuous excess near $(RA,Dec)=(270.8,-25)$. Tracing back the
position of these ``excess'' stars in the defining color-color diagrams
$(r-i,i-J)$ or $(r-i,i-H)$ we checked that they lie just above the
threshold, at moderately large reddenings.
The most plausible explanation for the absence of a similar clustering
among the $M_{gri}$ stars in panel~$a$ is that along that line of sight
the NIR reddening law is slightly steeper than elsewhere, in the sense that
stars deviate above the main trends in the relevant color-color
diagrams. There is no indication of any type that a real cluster of
stars exists around that position, from either CTTS diagnostics or Gaia
data. The possibility of a nonuniform reddening law should be kept in
mind before interpreting too naively the results from the M-star
selection method. Under this respect, the three different ways of
selecting M stars are useful for a mutual plausibility check, as in this
case.

The distribution of M stars in the cluster (inside the Chandra FOVs and
in their immediate neighborhood) is remarkably consistent between the
$M_{gri}$, $M_{riJ}$, and $M_{riH}$ subsamples (panels~$a$, $b$, and $c$).
It also agrees very well with the distributions of \ha-emission stars,
NIR- and UV-excess sources in panel $d$. Just to the West of the
rightmost Chandra FOV, the M-star selection finds small groups of
candidate members, which are also seen in panel~$d$: these are among the
westernmost stars for which there is convincing evidence of membership
to \ngc. Kalari \e (2015) also remarked that these stars are good
candidate members on the basis of \ha\ and UV excess, and their
neighborhood should be more carefully studied (the field examined in that
work is outlined in panel~$d$). In their vicinity lie the stars
7~Sgr (F2II-III) and HD~164536 (O7.5V), also indicated in panel~$d$,
which were found probably responsible for the local expansion of the
ionized nebular gas by Damiani \e (2017a).

Near the southern edge of the Kalari \e (2015) field, approximately
15~arcmin south of the cluster core, a small group of \ha\ emission
stars is found in panel~$d$, whose membership to \ngc\ needs to be considered 
carefully. However, we do not find them among the M-type candidates, nor
among the Gaia candidates (see below), so it becomes more likely that
they are an unrelated group of emission-line stars. As already shown in
Fig.~\ref{smoothed-halpha}, \ngc\ lies $\sim 1.5^{\circ}$ below the
Galactic Plane, which lies to the northwest of our studied field. The
increase in the density of NIR-excess objects towards the northwest, as
seen also in panel~$d$, is therefore ascribable to a diffuse galactic
population, not connected to \ngc.

To the southeast of the \ngc\ core, partially inside the eastern Chandra
FOV, a group of candidate members is found (from both the M-star
criteria and the various CTTS criteria), which follows the profile of
the local bright-rimmed cloud, in which the massive protostar M8-East IR
(Wright \e 1977)
is embedded. A group of hard (probably embedded) X-ray sources close to
that bright rim was also noted in the first Chandra X-ray study by Damiani
\e (2004).

Near the Hourglass nebula (center of rightmost Chandra FOV) no M-type
members are found: this is a bias, caused by the lack of data in the
VPHAS+ catalog from that neighborhood in the $r$ band (probably as a
result of the very bright \ha\ nebulosity), which is required
by all three M-star selection criteria.

\begin{figure}
\resizebox{\hsize}{!}{
\includegraphics[angle=90]{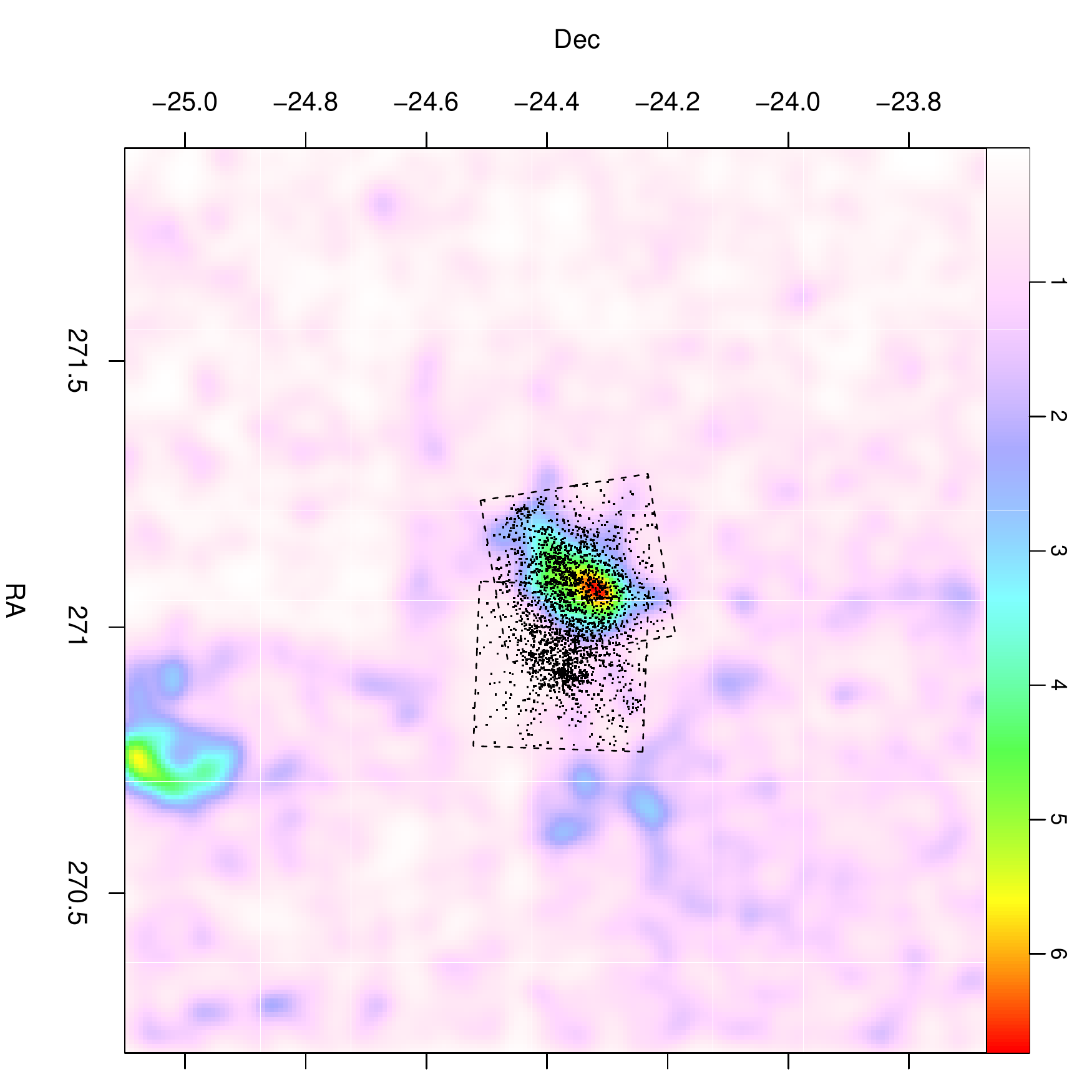}}
\caption{
Smoothed spatial distribution of all low-reddening M stars.
The black dashed regions are the Chandra FOVs, and dots are X-ray sources.
Top-axis colorbar as in Fig.~\ref{smoothed-halpha}.
\label{smoothed-Mstars}}
\end{figure}

Figure~\ref{smoothed-Mstars} is a slightly smoothed map of the
spatial distribution of all low-reddening M-type stars, compared to that
of individual Chandra X-ray sources. We again remark that the rightmost
Chandra FOV is much deeper than the other one, so that X-ray detections
there are overrepresented. In the same FOV, the lack of M-type stars
just discussed is evident. In the cluster core (left) FOV, instead, we
find a very good agreement between the distribution of X-ray sources and
that of M stars, indicating once again that the vast majority of both samples
consists of very good candidate members of \ngc. The southeastern region
(partially inside the lower-left corner of the left Chandra FOV)
has a distinct morphology, slightly separated from cluster core and
elongated along the bright rim (see e.g.\ the VPHAS+ \ha\ images in
Kalari \e 2015 or Damiani \e 2017a); the M-star sample shows that the
member distribution along the rim continues even slightly outside the Chandra
FOV to the East.

To the West of the Chandra FOVs, three groups of M stars are clearly
found, of which two associated with the stars 7~Sgr and HD~164536 as discussed
above, and a third (rightmost) one not associated with anything obvious.
In the next subsections we will discuss that this latter subgroup is
likely not associated with the cluster.

\begin{figure*}
\resizebox{\hsize}{!}{
\includegraphics[]{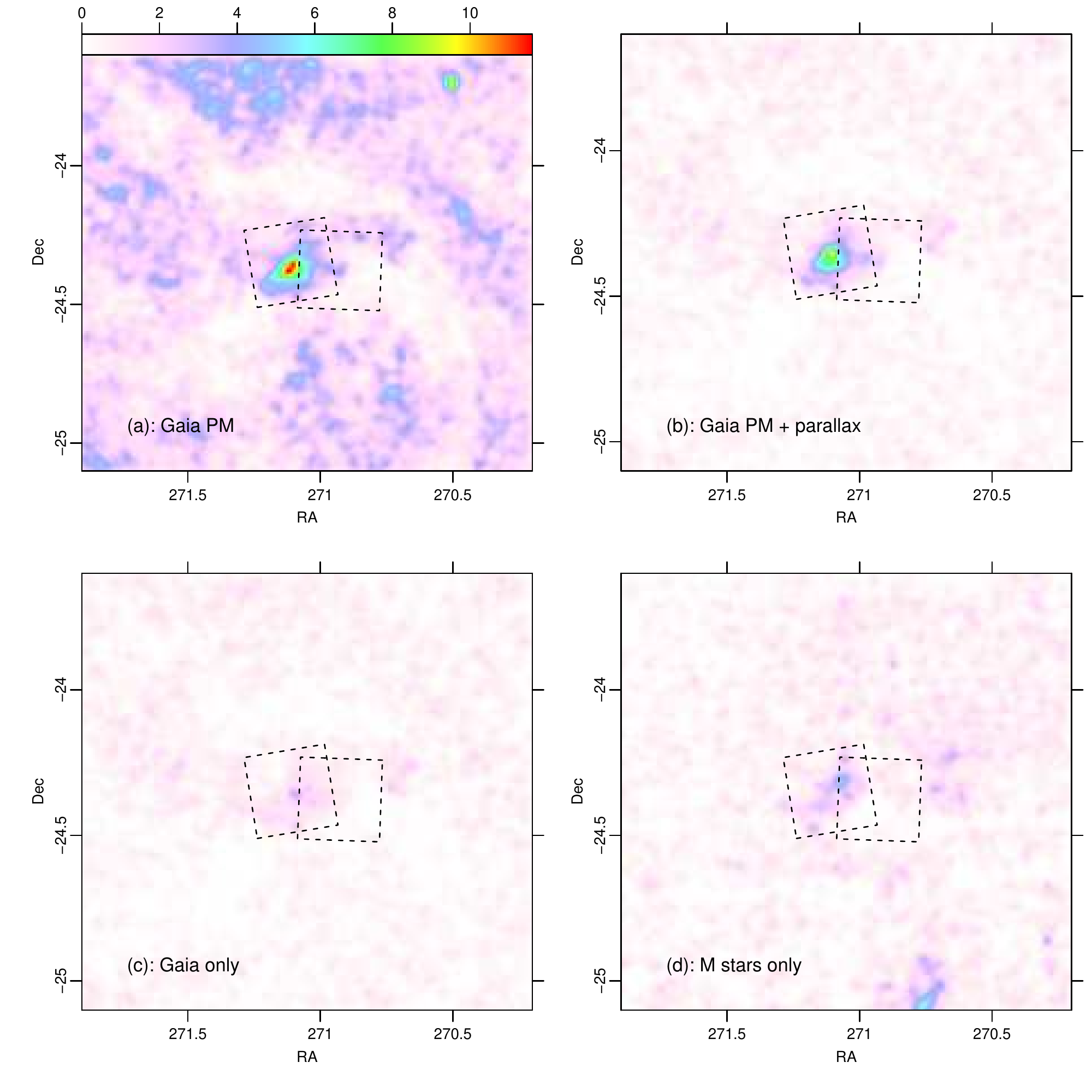}}
\caption{
Spatial distributions of subsamples selected according to various
criteria. Dashed squares are the Chandra FOVs. The color scale is the
same for all panels.
Panel $(a)$: All Gaia sources selected from proper motion alone, falling
inside the rectangle in Fig.~\ref{gaia-all}$b$.
and errors on $\mu_{\alpha}$ and $\mu_{\alpha}$ less than 0.5.
Top-axis colorbar as in Fig.~\ref{smoothed-halpha}.
Panel $(b)$: Gaia sources selected from proper motion and parallax,
falling inside both rectangles of Fig.~\ref{gaia-all}$b$ and $c$.
Panel $(c)$: Gaia sources as in panel~$b$, but missed by the X-ray, CTTS,
and M-star selection methods.
Panel $(d)$: M stars missed by the Gaia, X-ray, and CTTS selection methods.
\label{gaia-spatial-diff}}
\end{figure*}

Next, we have compared the spatial distributions of candidate members
according to different combinations of selection criteria, to study the
respective performances. For example, Fig.~\ref{gaia-spatial-diff}$a$
shows the result of selecting stars exclusively using Gaia proper
motions (ignoring parallaxes) in the cluster box of Fig.~\ref{gaia-all}:
this is not very effective, the \ngc\ cluster being still recognizable,
but accompanied by several other
local density enhancements. After adding the parallax constrain
(panel~$b$), instead, \ngc\ becomes much
better defined, together with its small peripheral subgroups of likely
members mentioned above, such as those in the southeast, and two out of
three to the West of the Chandra FOVs. Since all these structures were
already found by other selection criteria, the next question is whether
the Gaia data do actually add any new members to the list assembled from
M stars, CTTSs, and X-rays: panel~$c$ shows candidate members exclusive
to the Gaia sample, and it does indeed prove that Gaia still adds some
members to the pre-existing list, with a comparable distribution in space
(including subgroups). The complementary question was if the M-star
methods is of any usefulness, once we have assembled the Gaia, X-ray,
and CTTS sample: panel~$d$ shows that there are still M stars which
failed all other selection criteria, clearly with the same spatial
shape of the cluster. We conclude that all four methods we used here
are highly complementary to one another, each one being most effective
in a particular region of the (large) parameter space covered by the
\ngc\ member stars. Therefore, the \ngc\ member list obtained from the
combination of all of these criteria is definitely more complete than
those obtained from the individual criteria, or even by omitting just
one of them.
{ The exact completeness limit at the low-mass end depends on age and
extinction, as can be seen from Fig.~\ref{max-distance}.}
We should remark that this holds in this particular cluster,
whose members span a wide range of masses, extinction values, ages, and
other peculiarities typical of the PMS phase; we expect that a similar
situation holds for other { very young} clusters, but not for older clusters
containing only main-sequence stars (e.g., the Pleiades) or giants.

\begin{figure}
\resizebox{\hsize}{!}{
\includegraphics[]{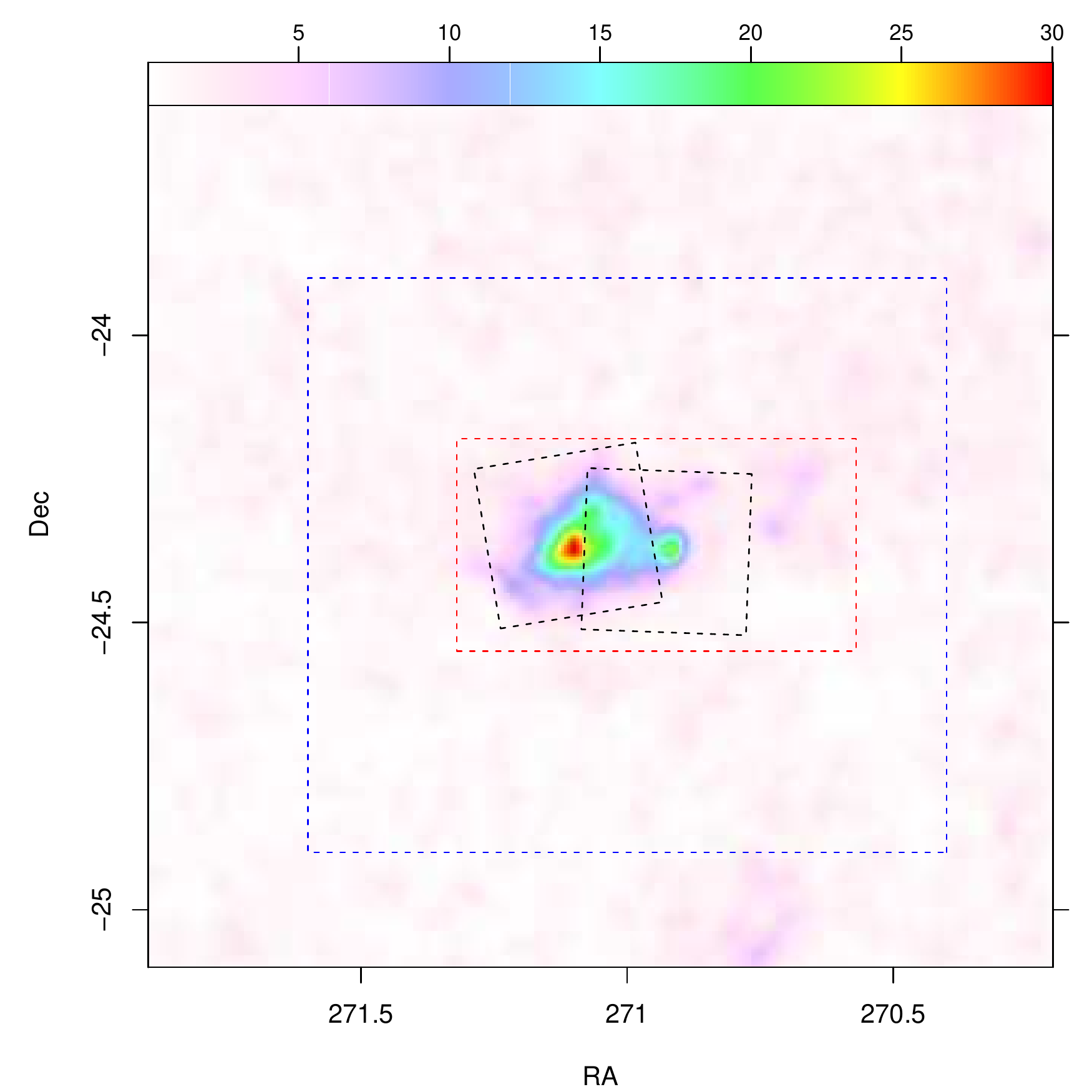}}
\caption{
Spatial map of candidate members selected by any method: Gaia proper motion
and parallax, CTTS diagnostics, M-type selection (but brighter than the
red line in Fig.~\ref{cmd-gaia-sel}), and X-rays.
The black squares are the Chandra FOVs.
The inner (red) rectangle is the final ``cluster'' region, while the outer
(blue) one is the same reference region as in Fig.~\ref{smoothed-halpha}.
\label{spatial-all}}
\end{figure}

\begin{figure*}
\resizebox{\hsize}{!}{
\includegraphics[]{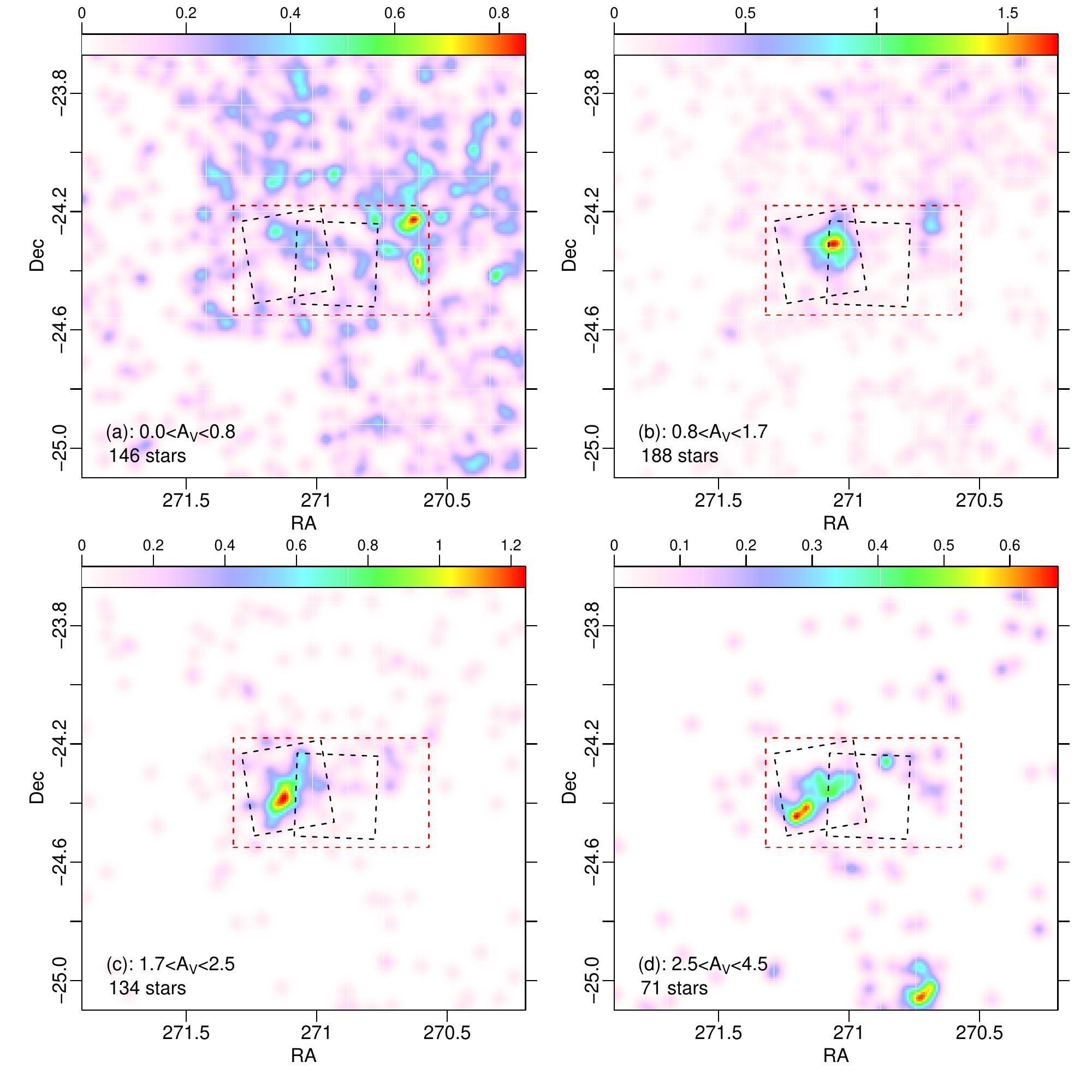}}
\caption{
Maps of spatial distribution of low-reddening M stars, for different
$A_V$ ranges. For $A_V>4.5$~mag (not shown), there is no sign of
clustering around the \ngc\ position. Dashed squares are the Chandra FOVs.
{
The dashed red rectangle is the cluster region as in
Fig.~\ref{spatial-all}. The number of M stars in this region is
indicated for each $A_V$ range.
}
\label{smoothed-Mstars-av}}
\end{figure*}

Finally, the complete member list obtained from combining all our
selection methods yields the spatial distribution shown in
Fig.~\ref{spatial-all}. The comparison between the cluster-related
density enhancement with the fluctuations in the surrounding field-star
density suggest strongly that all possible cluster members are contained
within the rectangular regions shown in red: this is our ``final''
cluster region. There is no diffuse population of members outside of it.
The secondary peak at the Hourglass, where optical data are highly
incomplete because of the very bright nebulosity, is clearly defined
thanks to the X-ray data. Since there is no comparably bright
nebulosity elsewhere in the studied region, it is unlikely that the
optical data (including Gaia) would miss a substantial percentage of stars
in any subregion other than the Hourglass.
The southeastern M8-East region (along the bright
rim) is also identifiable in Fig.~\ref{spatial-all} with its elongated shape,
as are also the minor condensations to the northwest.

\subsection{Extinction}
\label{extinct}

One of the very useful properties of M stars is that their position on
the color-color diagrams of Fig.~\ref{definitions}-$a,b,c$ is not
degenerate between $T_{eff}$ and extinction $A_V$. Therefore, adopting a
zero-extinction locus (from a theoretical isochrone) and a reddening
vector we are able to determine $A_V$ for each individual M star. In
particular, we use $A_V$ as determined from the $(r-i,i-J)$ diagram,
which yielded the largest number of candidates.
{
In this way we have derived $A_V$ for 3615 M stars in the entire region, of
which 627 in the cluster region (55\% of all cluster M stars).
}
We therefore show the spatial distributions of M stars in four different
$A_V$ ranges in Fig.~\ref{smoothed-Mstars-av}. The literature $A_V$ for
\ngc\ is 1.08. In panel~$a$, M stars with $0<A_V<0.8$ do not show any
clustering at the cluster position, and it seems legitimate to conclude
that nearly all stars in this extinction range are foreground field stars.
Two of the subgroups to the West of the Chandra FOVs are clearly
found in this extinction range, however.

\begin{figure}
\resizebox{\hsize}{!}{
\includegraphics[angle=90]{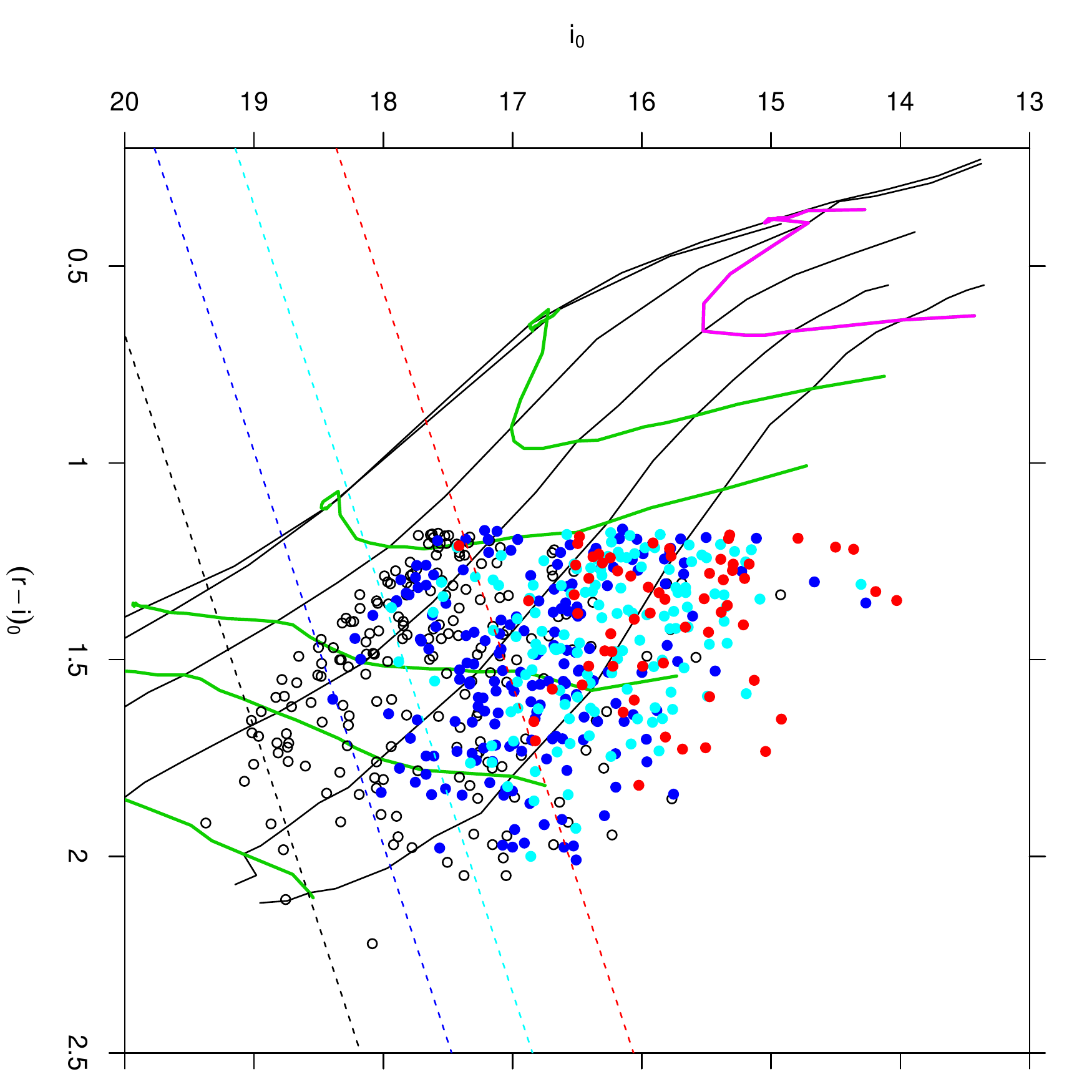}}
\caption{
Dereddened CMD for M stars in final ``cluster'' region defined in
Fig.~\ref{spatial-all}.
Evolutionary tracks and isochrones as in the same figure, but with zero
reddening.
Colored dots are individually-dereddened M stars, with color-coded
extinction ranges. Black: $0.0<A_V<0.8$; Blue: $0.8<A_V<1.7$;
Cyan: $1.7<A_V<2.5$; Red: $2.5<A_V<3.5$.
Oblique dashed lines indicate completeness limits for the same $A_V$ ranges.
\label{cmd-dereddened}}
\end{figure}

In all higher extinction ranges, instead, the cluster is very evident,
but the higher-extinction members are distributed differently from the
low-extinction members. The third (westernmost) subgroup outside the
Chandra FOVs, mentioned above, disappears at higher extinction, while
the two closer ones (associated with 7~Sgr and HD~164536) are still
visible, and therefore partially obscured by the M8 dark nebulosity. Not
surprisingly, the highest extinction M-type members (panel~$d$) are
mostly found along the southern bright rim mentioned above, and
partially embedded in its dust. Less expected, instead, is the presence
of high-extinction members near the cluster core, where most stars have
low extinction. Stars in that direction, therefore, are distributed
along the line of sight at different distances, even until fairly deep
within the dust cloud. In panel~$d$, a bright spot near the top of the
rightmost Chandra FOV coincides with a compact group of Chandra X-ray
sources, also visible in Fig.~\ref{smoothed-Mstars}, which therefore
constitute likely a small group of very young, embedded low-mass stars.
There is no indication of a diffuse population of \ngc\ members, in any
$A_V$ range.
{
Overall, the largest number of clustered M stars falls within the $A_V$
range 0.8-2.4, with a tail ($\sim 20$\% of the stars) at higher $A_V$.
}

While interpreting this type of maps, however, it should be kept in mind
that larger extinction values correspond to decreased completeness in
the sample. The extinction corrected CMD of M stars in the cluster
region is shown in Fig.~\ref{cmd-dereddened} for stars in four
extinction ranges, where we also show the completeness limit at the end
of each range (i.e., we show conservative estimates for the limits).
Stars in the lowest extinction range are not expected to be members, as
just discussed. At higher extinction, the lowest-mass stars can only
be present in the lowest $A_V$ range, and significantly less present for
increased extinction. The same holds for the oldest cluster members.
This effect would introduce a spurious correlation between age and
extinction if not properly accounted for, as that one would naively infer
from the same diagram, by noticing that blue dots (less extincted stars)
lie systematically below the red dots (highest-extinction stars).
{ The existence of still higher-extinction \ngc\ members, which are missing
in this and other studies based on optical data, is proven by the detection of
64 Class~0/I IR sources by Kumar and Anandarao (2010).}

\begin{figure}
\resizebox{\hsize}{!}{
\includegraphics[]{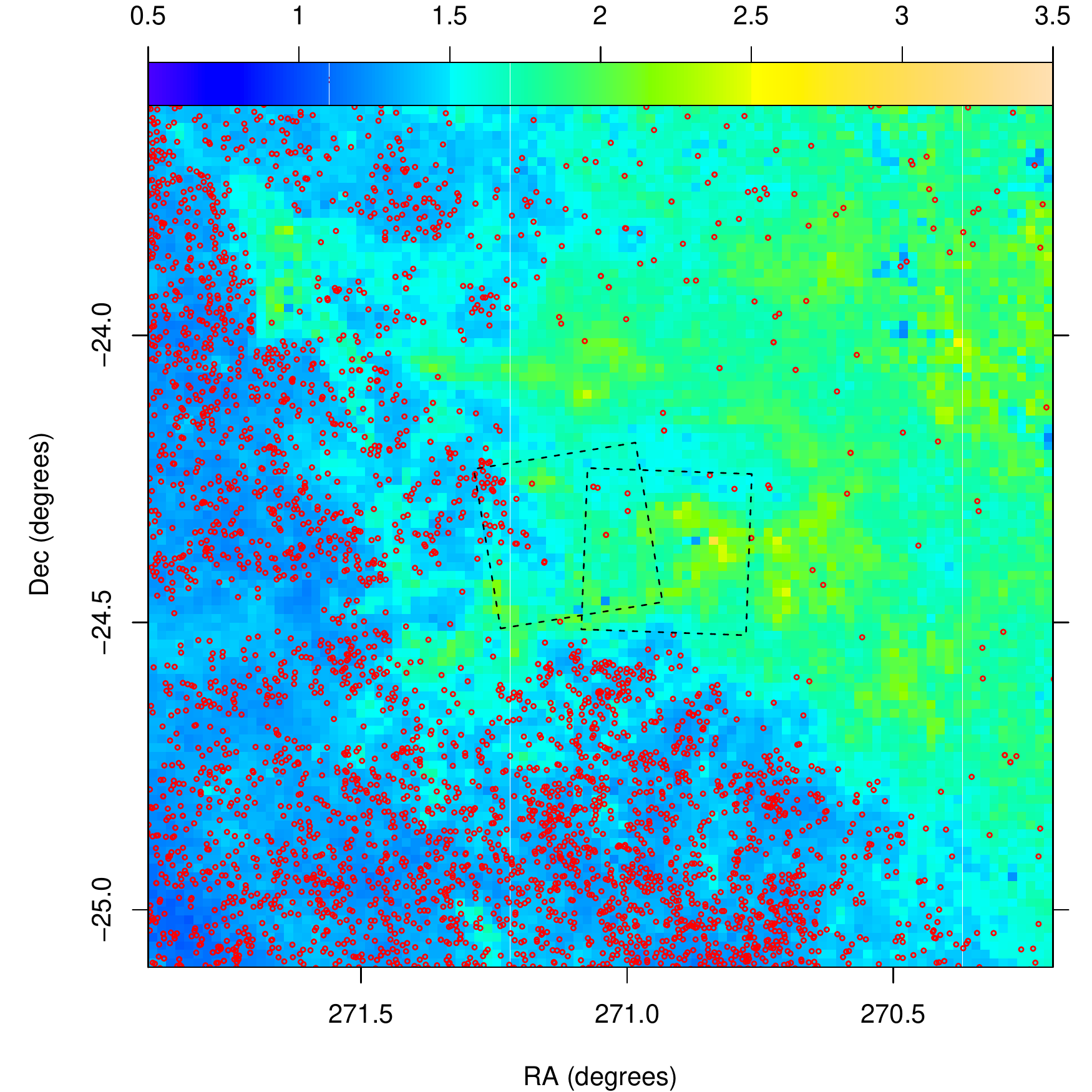}}
\caption{
Map of mean $(J-H)$ color (color scale shown in top axis).
Red circles are high-reddening M stars (giants) from
Fig.~\ref{definitions}-$a,b,c$.  Dashed squares are the Chandra FOVs.
\label{extinction-map}}
\end{figure}

We have also studied the spatial distribution of extinction behind the
cluster, averaged over the line of sight until large distances.
This help in understanding the environment in which \ngc\ is found.
The average extinction is well approximated by the average $J-H$ color,
computed in small spatial bins. The same approach was successfully taken
in Paper~I, and earlier in Damiani \e (2016). The map of average $J-H$
is shown in Fig.~\ref{extinction-map}, from which one may infer
extinction, being $A_V \sim 9.5 \times (J-H-0.5)$. There is a very good
agreement between the zones with the largest extinction from this map,
and those with lowest source density from the Gaia density map of
Fig.~\ref{gaia-spatial}. The regions close to the Hourglass are
confirmed to be among the most extincted sightlines towards \ngc\
{ and the Lagoon Nebula.} Behind
the cluster core there is only a moderate amount of extinction, compared
to the Hourglass, and also to the outer regions of the nebula, which
appear to bifurcate to the North and South of the cluster core (and the
Chandra FOVs). The region southeast of core, near the bright rim where
also a number of cluster members are found, is also one of large
extinction. Otherwise, the eastern part of the cluster (farthest from
the Galactic plane) does not seem to be bound by a thick dust wall,
but relatively clear, and the overall morphology of dust suggests that
the cluster lies in a cavity carved inside a thick dust envelope, which
has already broken on the side opposite with respect to the Galactic
plane: these were likely the lowest-density parts of the original dust
cloud, and the first ones which were dissipated.

Already very close to the cluster core, the line-of-sight extinction
drops to low enough values that distant giants begin to be
detectable (red dots in the figure). \ngc\ and { the Lagoon Nebula are}
projected onto the
Galactic Bulge, being only $\sim 6^{\circ}$ away from the Galactic Center.
Giants are found in large numbers only in the regions with relatively lower
densities (i.e., not exceeding 10-15 mag in $V$), in agreement with
their $A_V$ values that can be derived from the diagrams in
Fig.~\ref{definitions}-$a,b,c$; their spatial distribution matches the
distribution of extinction very well, providing us with a very coherent
picture. Especially the southeastern parts of \ngc\ are surrounded by
lower-density sightlines. Therefore, the corresponding obscuring dust is
likely a protruding cloud, with very little dense material all around.
Lacking pressure balance against the (radiative and mechanical) push
from the massive stars in \ngc, the southeastern part of the cloud will
probably disperse soon after it ceases to form new stars.

\subsection{Ages of members}
\label{age}

\begin{figure*}
\resizebox{\hsize}{!}{
\includegraphics[]{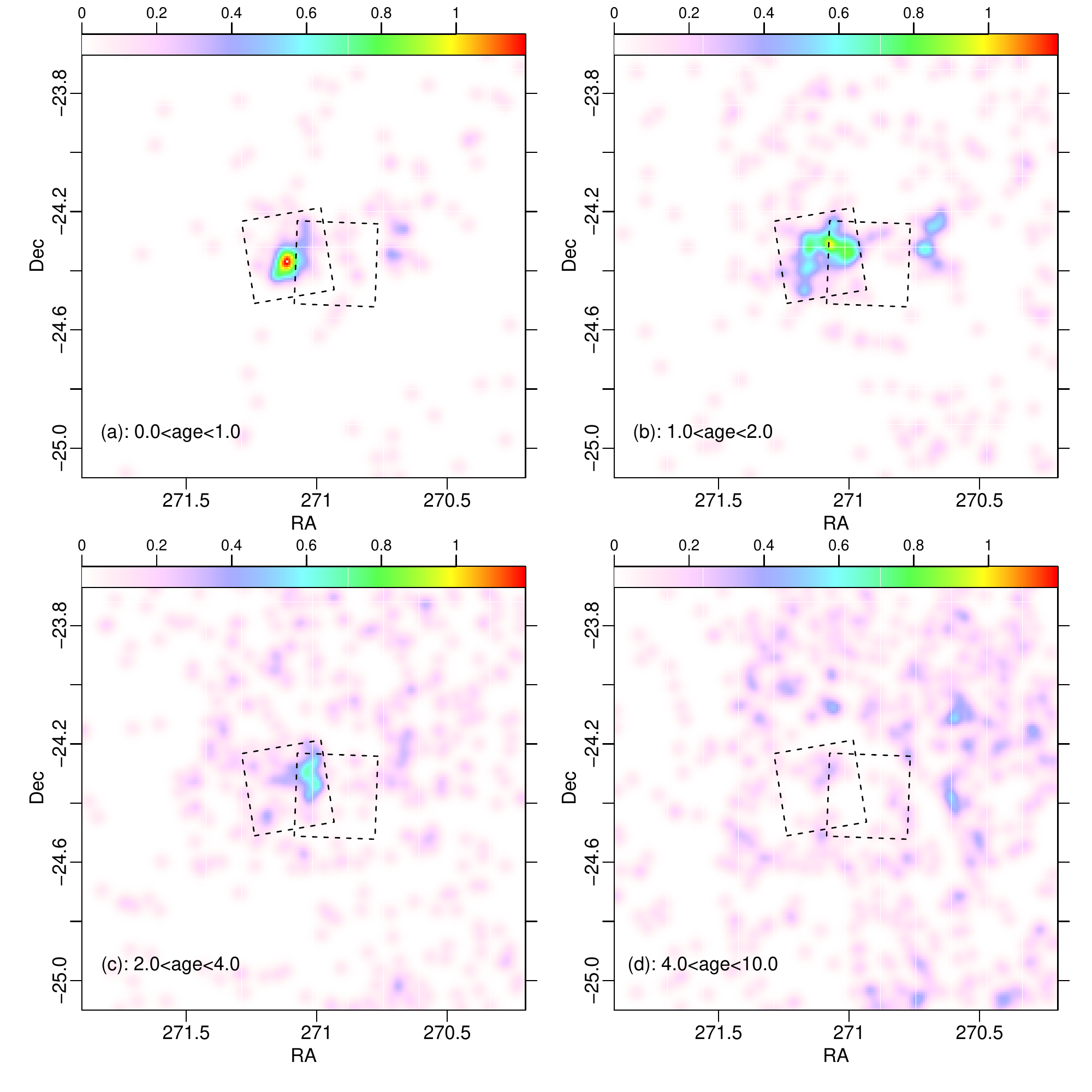}}
\caption{
Spatial distributions of M stars in different age ranges, as indicated
(in Myr).
\label{spatial-ages}}
\end{figure*}

After placing M stars individually on the extinction-corrected CMD, we may
determine more accurately their ages by comparison with BHAC isochrones.
These age values are only valid for stars at the assumed distance (no
selection on parallax is made here). We therefore obtain some
indications on the star-formation history in { the Lagoon Nebula}
from the spatial
distributions of M-type members in several age ranges, shown in
Fig.~\ref{spatial-ages}.
Very few, if any, cluster members are found to be older than 4~Myr in
panel~$d$. This
statement must be accompanied by the caveat that we are only able to
find the most massive M-type stars in this age range, as also predicted
by the MDA diagrams of Fig.~\ref{max-distance}. For younger ages our
results are more reliable (panels~$a-c$): star formation appears to have
begun in the region common to the two Chandra FOVs, close to the most
massive cluster member (the O4 star 9~Sgr); then, it both continued in
the same place, and propagated to the East; afterwards, star formation near 
to 9~Sgr stopped, while it became more active to the East. We are
unfortunately unable to find M stars in the Hourglass region using the
VPHAS+ data, although stars embedded in the Hourglass nebula itself
are likely to be very young. Stars in the western subgroups appear to
have intermediate ages.

Such a complex pattern of formation events with time and space might
justify why different authors have come to different conclusions on the
formation history in \ngc, on the basis of different studied samples,
and therefore under different selection biases (see e.g.\ Kalari 2015).
Also in the present work, the extinction-related bias as discussed above,
as well as the lack of suitable optical photometry near the Hourglass,
prevent to obtain a definitive picture of the star-formation
history { in the Lagoon Nebula}.

\begin{figure}
\resizebox{\hsize}{!}{
\includegraphics[]{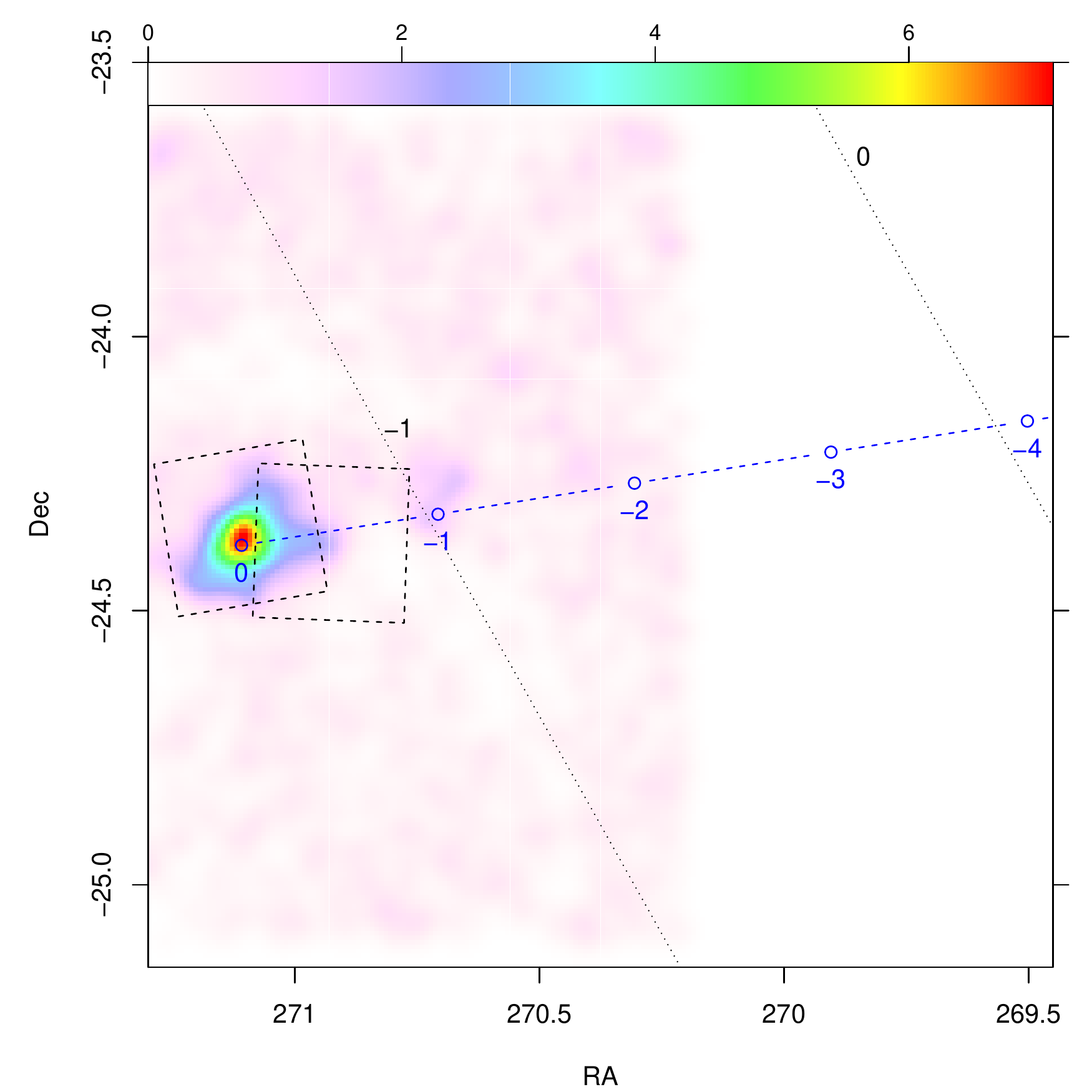}}
\caption{
Smoothed spatial distribution of Gaia members, plus back-projected
mean cluster position, from about 4~Myr ago to now (blue dot-dashed
line). Blue labels indicate time in Myr.
Dotted black lines indicate constant Galactic latitude $b$, as labeled.
\label{back-project}}
\end{figure}

\subsection{Cluster dynamics}
\label{dynamics}

We have presented in Sect.~\ref{plx} the proper-motion distributions of
both the cluster and the surrounding field stars.
We here consider the average cluster proper motion, in an attempt
to estimate the path followed by the cluster in the last few Myrs.
The median proper motion of the X-ray detected Gaia members in the
cluster region (a highly reliable sample) is $(\mu_{\alpha},
\mu_{\delta}) = (1.268, -1.996)$~mas/yr. Therefore, the relative motion
with respect to the average Galactic-plane population in the same
direction is
$(\Delta \mu_{\alpha}, \Delta \mu_{\delta}) = (1.520, -0.201)$~mas/yr.
This relative motion implies that the cluster has travelled a
significant distance in the last few Myrs. Fig.~\ref{back-project} shows
this path, and demonstrates that \ngc\ is likely to have crossed the
Galactic plane { around 4~Myr} ago. It is intriguing that this is
approximately the age of the oldest members whose age we can measure.
A very plausible scenario is therefore that the encounter between the \ngc\
parent molecular cloud, coming from above the Plane, and the Galactic
plane itself, did generate a rapid and strong compression of the cloud,
leading to rapid star formation while the cloud continued its travel towards
negative Galactic latitudes.
{
Collisions between molecular clouds and the Galactic disk were modeled by
Tenorio-Tagle \e (1987) and Comeron and Torra (1992), and were suggested
to lead to triggered star formation by (for example) Comer\`on (2001).
}

\begin{figure*}
\resizebox{\hsize}{!}{
\includegraphics[angle=90]{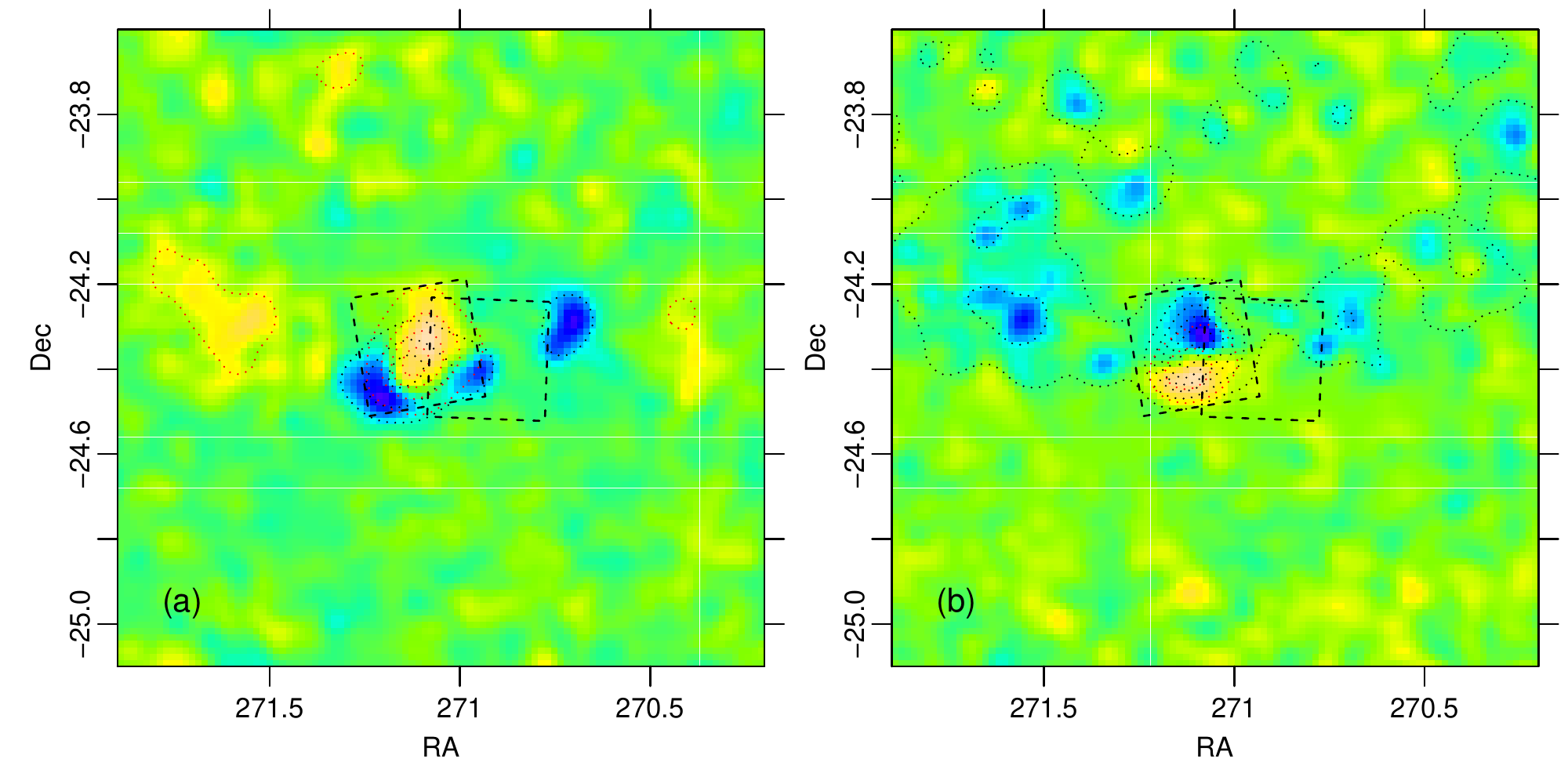}}
\caption{
$(a)$: Difference between spatial distributions of Gaia members (according to
both proper motion and parallax) with $\mu_{\alpha}$ less and above
1.33~mas/yr, respectively. Only Gaia sources with $\mu_{\alpha}$ errors
less than 0.5~mas/yr are considered.
In blue (yellow)
are negative (positive) values, with green indicating zero difference.
Dashed squares are the Chandra FOVs. The red and black dotted contours
describe the original distributions in the two $\mu_{\alpha}$ ranges,
whose difference is shown by the map.
$(b$): Same as $a$, but showing the difference between distributions of
stars with $\mu_{\delta}$ less and above -2.0~mas/yr.
\label{gaia-pm-correl}}
\end{figure*}

For the same member sample, the observed ranges of $\mu_{\alpha}$ and
$\mu_{\delta}$ (2.5~mas/yr for both) are about 10 times larger than the
median $\mu_{\alpha},\mu_{\delta}$ errors. This opens up the possibility
that the internal cluster dynamics might be resolved with these data.
We have therefore computed the spatial distributions of $\mu_{\alpha}$
for Gaia member stars, separately for $\mu_{\alpha}$ less and above
1.33~mas/yr, and show their (normalized) difference in
Fig.~\ref{gaia-pm-correl}$a$. Here yellow indicates a relative
preponderance of stars moving towards right (lower $RA$) with respect to
the mean cluster motion, while blue indicate stars in relative motion
towards left. The same was done for $\mu_{\delta}$ (above and below a threshold
of -2.0~mas/yr, see Fig.~\ref{gaia-pm-correl}$b$). Also in this latter case, 
yellow indicate stars in relative motion towards negative $Dec$, and
blue the opposite. We observe that there is near symmetry of these relative
motions with respect to the cluster core (the center of the
left-hand Chandra FOV), and the sense of motion indicates expansion away
from this center, along both $RA$ and $Dec$. Near the cluster periphery
different motion patterns are instead found, both close to the Hourglass
and in the western group, which both move towards East. It should be
remarked however, that such motions are averages with non-negligible
internal spread (due to both true random motion and measurement errors),
as seen from the individual distributions used to
produce the difference images in Fig.~\ref{gaia-pm-correl}; these are
also shown in the same figure by red and blue contours, which largely
overlap.
{ Similar results on expansion in \ngc\ have recently been obtained
by Kuhn \e (2018) and Wright \e (submitted).}

\begin{figure}
\resizebox{\hsize}{!}{
\includegraphics[]{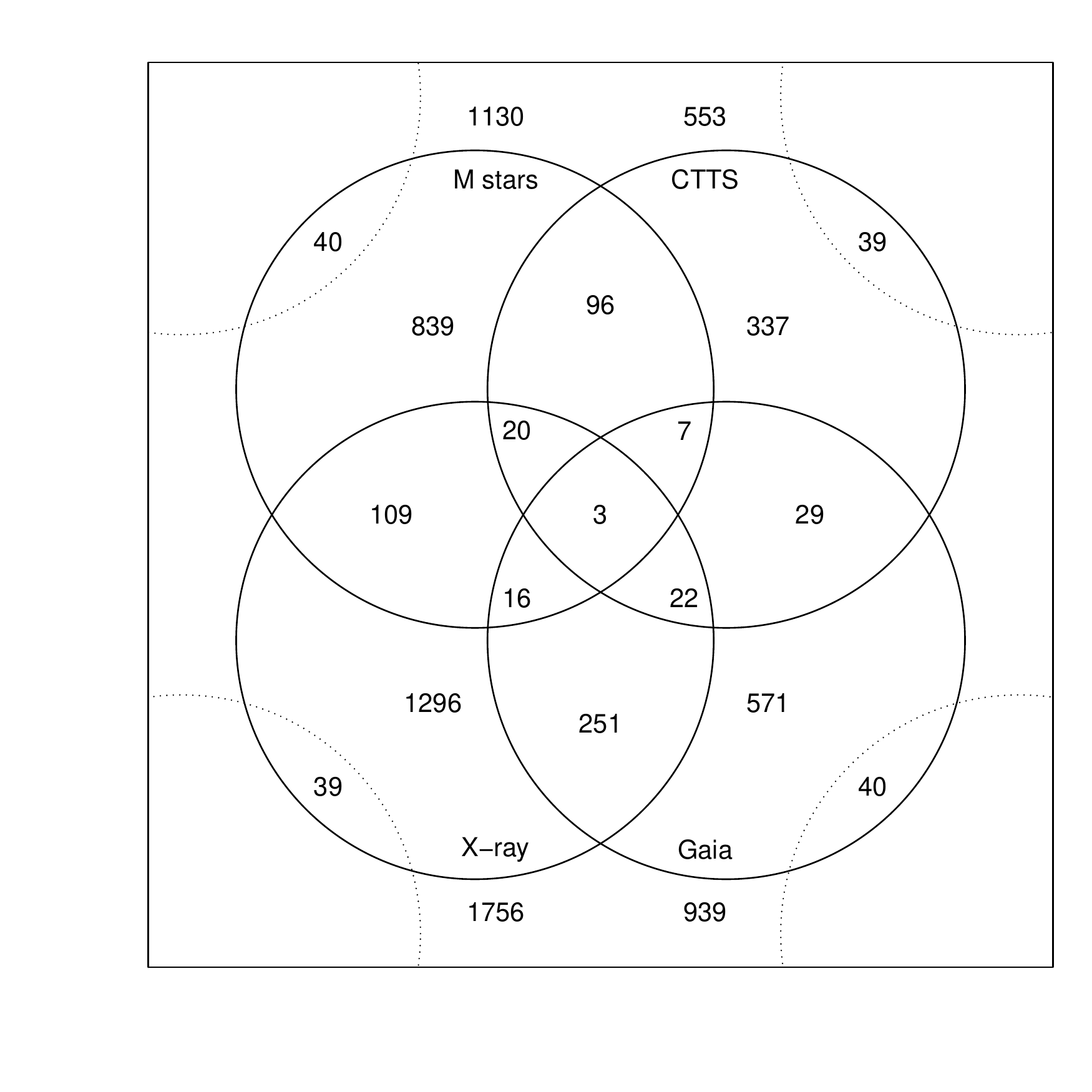}}
\caption{
Venn diagram showing the intersections among the different member
subsamples. The total number of candidate members for each subsample is
shown outside the corresponding circle, near its label.
The diagram wraps around its
corners, in order to show the exclusive intersections between X-ray and
CTTS samples, and between Gaia and M-star samples.
\label{venn}}
\end{figure}

\subsection{Member statistics and contamination}
\label{stat}

Results from the previous section have shown that the different member
selection methods used are mutually complementary, each one being most
effective on some parts of the large parameter space (range of masses,
ages, extinction values, importance of disk- and accretion-related
phenomena) occupied by \ngc\
members. In this section we examine this more quantitatively, and try
to estimate the level of contamination which affects subsamples found
by any particular method.

Figure~\ref{venn} is a Venn diagram showing the intersection between
the four selection criteria employed. For simplicity, we have put in a
single sample (CTTS) all stars selected by either \ha\ emission, NIR- or
UV-excesses. All samples are spatially limited to the final cluster region.
The M-star and CTTS samples include the photometric threshold of Sect.~\ref{photlim}.
The ordinary Venn diagram applied to four samples with such a complex
pattern of overlaps does not permit to show graphically all pairwise
intersections: this is the case of the diagonally opposite samples
(e.g., X-ray and CTTS samples), whose exclusive intersection (i.e.,
with no stars in common to the Gaia and M-star samples) cannot be represented
graphically. Therefore, we have ``folded'' around its corners the diagram
to show such intersections, so that, for example, the lower-left dotted circle
duplicates the CTTS sample in the upper-right solid circle,
and its exclusive intersection with the X-ray sample
({ 39} objects) is made visualizable. The same holds of course for all four
samples. By symmetry, the X-ray-CTTS intersection is duplicated in the
upper-right corner, and the number of objects shown there refer to exactly
the same objects in the opposite corner: when computing a grand total
({ 3675} stars),
those objects must only be counted once. They are indicated twice in the
figure, however, to make the sum of numbers inside each circle equal to
the total number in the respective sample (indicated just outside each circle).
The complete photometry of the { 3675} candidate members is reported
in Table~\ref{table-members}, together with their classification as
optical/NIR excess stars, M stars, X-ray sources and Gaia members.

Stars fulfilling more membership criteria are better-confidence members.
However, the diagram shows that the number of stars fulfilling all criteria
is very small (three objects), so being extremely conservative in the selection would
give meaningless results. Also very small subsamples fulfill three criteria.
We therefore consider all stars with at least two criteria as reliable members
individually ({ 711} stars).
The largest overlap between two subsamples is that between
Gaia and X-rays, with 292 ($251+16+22+3$) common objects. Then,
M-stars and X-rays have 148 ($109+20+16+3$) stars in common,
M-stars and CTTS have 126 ($96+20+7+3$) stars in common, 
CTTS and Gaia have 61 ($22+7+3+29$) stars in common, 
Gaia and M-stars have 66 ($40+16+3+7$) stars in common, and
X-rays and CTTS have 84 ($39+20+3+22$) stars in common.
These relatively small numbers confirm the complementarity between
selection criteria.

\begin{figure}
\resizebox{\hsize}{!}{
\includegraphics[]{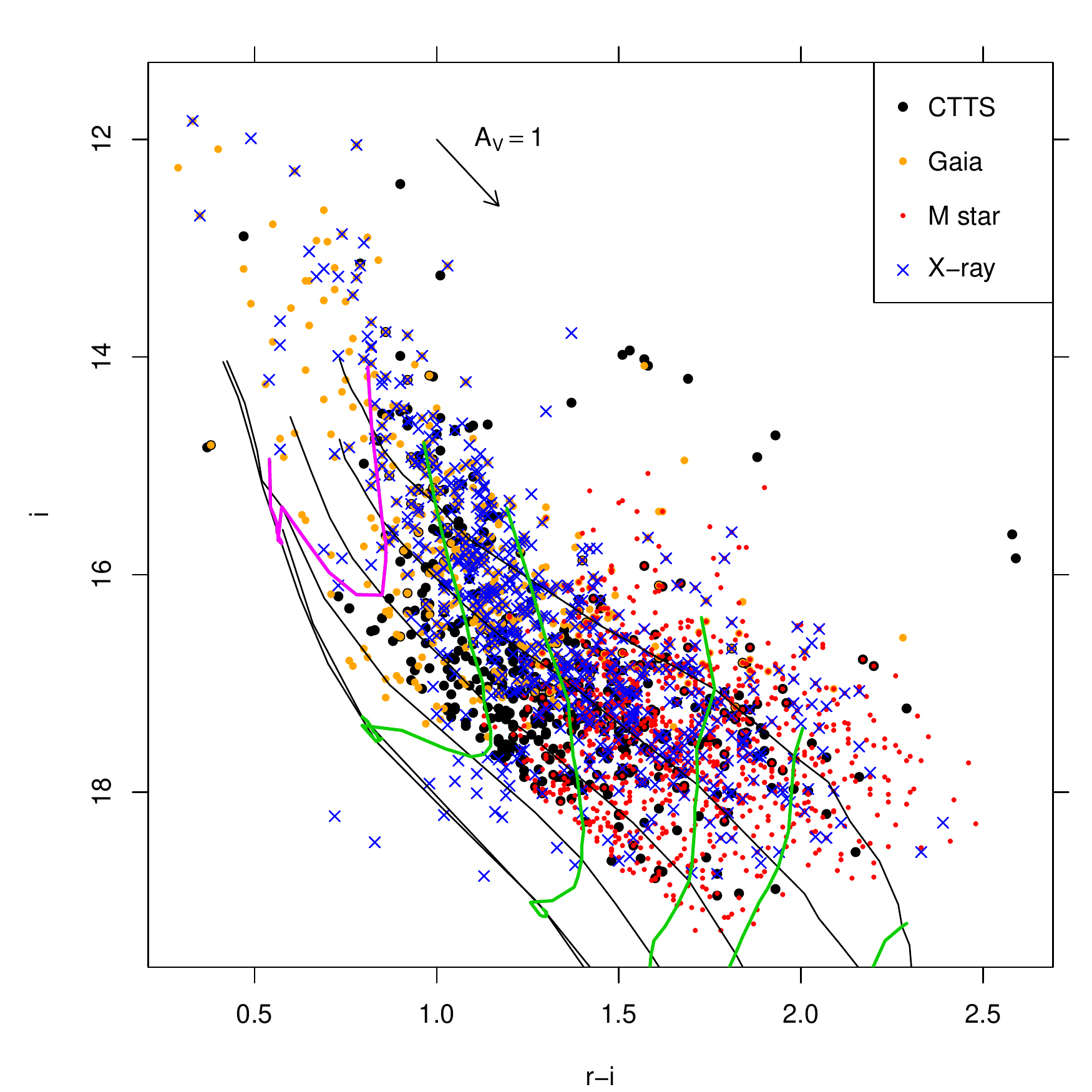}}
\caption{
CMD of final candidate members. Track and isochrones (ages of 1, 3, 10,
30, 100~Myr and 10~Gyr) as in Fig.~\ref{cmd-gaia-sel}.
\label{cmd-final}}
\end{figure}

Statistically, also stars with a single membership criterion
(i.e., 839 M~stars, { 337} CTTS stars, 1296 X-ray sources, and 571 Gaia
members, see Fig.~\ref{venn})
are members in a given percentage, as was qualitatively shown in
Fig.~\ref{gaia-spatial-diff}-$c,d$. The number of non-member contaminants
may be estimated by applying the same selection to the reference region
surrounding the final cluster region, and scaling for the respective sky areas.
The estimated contaminants are thus 525, { 89}, and 333 for the M-star,
CTTS, and Gaia samples, respectively (62.6\%, { 26.4\%}, and 58.3\% of the
total single-criteria samples, respectively).
Therefore, the estimated net number of members selected exclusively by
one criterion are 314, { 248}, and 238 for the M-star, 
CTTS, and Gaia criteria, respectively.
Unfortunately this cannot be done for the X-ray sample, there being no
Chandra data for the reference region, but in similar
X-ray pointings the number of field X-ray sources is of order of $\sim 100$.
These unrelated sources are however partly absorbed by those with
no optical-NIR counterpart, which we excluded from our analysis (see
sect.~\ref{xdata}), so that it is probably unnecessary to subtract them
here from the X-ray member count.
The total net number of \ngc\ members estimated with all our methods becomes
therefore of { 2728} stars, down to $0.2-0.4 M_{\circ}$, the exact
limit being dependent on individual stellar ages and extinction.
{
If we add the 671 X-ray detections without optical
counterpart, strongly clustered inside the Lagoon Nebula, we obtain a total
population of 3399 stars.

A CMD of all final members is shown in Fig.~\ref{cmd-final}. The same
qualitative considerations apply as for the CMD of Gaia-detected members
of Fig.~\ref{cmd-gaia-sel} (Sect.~\ref{photlim}). The M-star and CTTS
samples are most contaminated by field stars near the
lower envelope of datapoints (between the 10 and 30~Myr isochrones).
This is shown both by comparison with the CMD of the reference field
(not shown), and by the lack of clustering of stars at the oldest
apparent ages (Fig.~\ref{spatial-ages}). These stars are kept here for
completeness, but their individual membership needs a more careful
assessment.
It is noteworthy that, after considering all corrections for
contamination, the net percentages of CTTSs among all members
((553-89)/2728, or 17\%) or that among M stars (126/(1130-525), or 21\%)
are low compared to analogous percentages for other 1-2~Myr clusters (60-80\%
according to Haisch \e 2001). This might be related to disk
photoevaporation caused by the winds and radiation from massive stars,
as suggested by Guarcello \e (2007) for the young cluster NGC~6611.
}

\section{Conclusions}
\label{concl}

By combining various techniques for member selection, complementary to
each other, we have assembled a list of { 3675} candidate \ngc\ members
down to approximately $0.3 M_{\odot}$. Of them, only { 711} fulfill more
than one
membership criterion, but we estimate that the total net number of members
in our list is { slightly over 2700}.
Among the single-criterion subsamples, the
X-ray sample is estimated to be the least contaminated by field stars,
while contamination ranges from { 26\%} to 62\% for the M-star, CTTS and
Gaia selected subsamples.
Nevertheless, the statistical properties of the \ngc\ population are
very well defined, since even though the population in our different
subsamples do not coincide star-by-star, they often exhibit similar
general properties, which demonstrate the robustness of the results
obtained.

Results from our newly-developed method for selecting M-type cluster
members are validated by the new Gaia DR2 data, which in turn agree very
well with the X-ray selected members.

The Gaia distance to \ngc\ is found to be 1325~pc, which considering the
systematic error is compatible with our previous determination (1250~pc).
The \ngc\ morphology is confirmed to be complex, as from previous
studies, with a cluster core that contains the bulk of members, and
secondary concentrations of members in known regions (the Hourglass
nebula, the M8-East region) and anonymous, small groups (two groups in
the N-W part, close to the stars 7~Sgr and HD~164536). There is
no diffuse population of members all around the cluster.

Thanks to our new method for selecting M-type members, whose extinction
can be individually determined, we find that member stars lie behind a
variable amount of obscuration, with a defined spatial pattern. The
literature value of extinction ($A_V=1.08$, Sung \e 2000)
corresponds to the
lower bound to the $A_V$ range found here. The maximum extinction is not
well defined because of our limiting magnitude, but may occasionally be
very large (as in the case of the well-known embedded protostar M8-East IR).

Isochronal ages also show a complex spatial pattern, with a general
trends of age decreasing from
center towards (sky-projected) periphery.  There are young (and extincted)
stars even close to the (otherwise older and little obscured) cluster
core, suggesting that star formation has also proceeded along the line
of sight, towards deeper layers inside the molecular cloud.
The western regions are apparently not involved in this sequence, and
have an independent star-formation history from the rest of \ngc.
There is no significant population (if at all) of cluster stars older than
5~Myr.

The Gaia dynamical data show that the \ngc\ cluster and its parental
cloud crossed the Galactic plane about { 4~Myr} ago. Therefore, that
event might have been the main trigger of all star-formation phenomena
thereafter. The \ngc\ parent molecular cloud { (whose illuminated
part is the Lagoon Nebula)} is now travelling towards
regions of lower ambient density, and we find evidence that its border at the
lowest (negative) latitudes $b$ is being disrupted, probably by the
winds and radiation from the massive OB stars in the cluster.

The precision of Gaia proper-motion data also enables us to partially resolve
the internal cluster dynamics, and provide indications that the bulk of the
cluster is expanding away from its center. It is expected that Gaia data
from later data releases will permit even more detailed studies of the
\ngc\ internal dynamics. Also, deeper photometric surveys (e.g.\ from
LSST) will help discover the entire \ngc\ population down to the
substellar limit.

\begin{appendix}

\section{VVV-UKIDSS calibration}
\label{append1}

{
Here we examine the photometric differences between the King-UKIDSS and
VVV magnitudes in the $H$ and $K$ bands. Only sources with magnitude
errors less than 0.1~mag were used in each of the plots.
Fig.~\ref{vvv-ukidss-h} presents the comparison for the $H$ band and
Fig.~\ref{vvv-ukidss-k} for the $K$ band. The red dashed lines indicate
the average difference. No color dependence was found.

\begin{figure}
\resizebox{\hsize}{!}{
\includegraphics[angle=90]{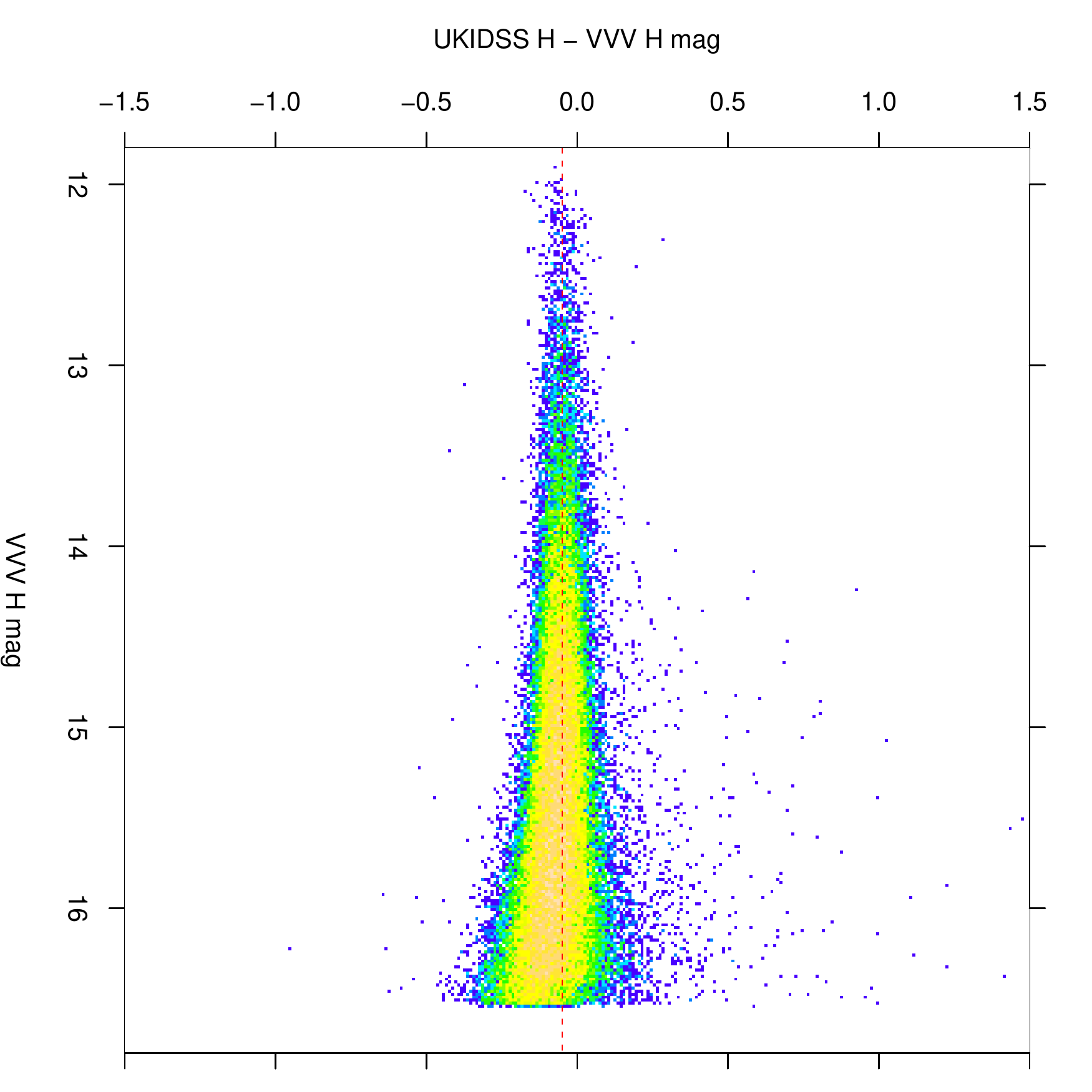}}
\caption{
Difference in $H$ magnitudes between King-UKIDSS and VVV catalogs, vs.\ VVV
$H$ magnitude. The red dashed line indicate the average difference of
-0.05 mag.
\label{vvv-ukidss-h}}
\end{figure}

\begin{figure}
\resizebox{\hsize}{!}{
\includegraphics[angle=90]{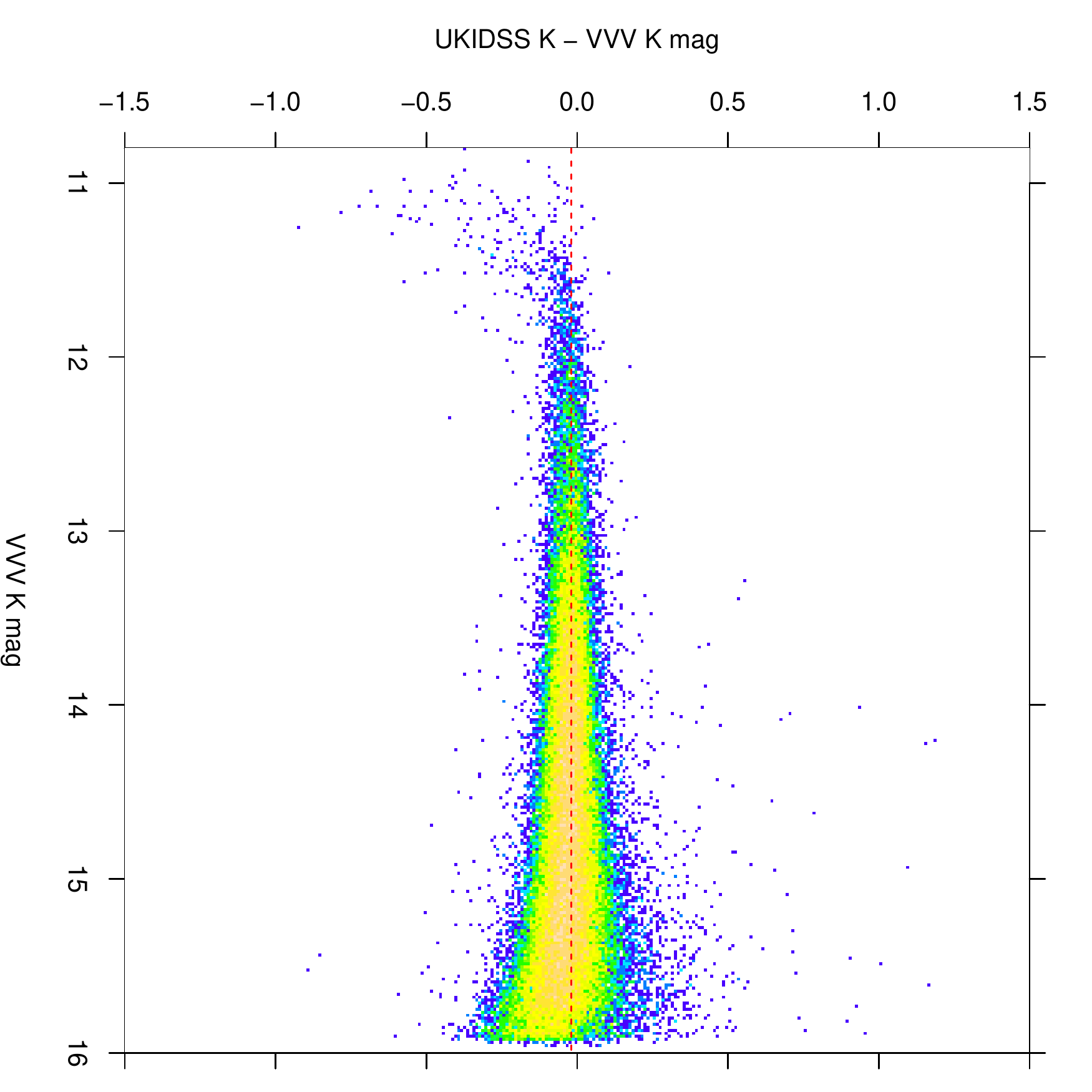}}
\caption{
Difference in $K$ magnitudes between King-UKIDSS and VVV catalogs, vs.\ VVV
$K$ magnitude. The red dashed line indicate the average difference of
-0.02 mag.
\label{vvv-ukidss-k}}
\end{figure}

\section{Reddening law}
\label{append2}

Here we provide some details on the reddening law used in this work,
which was as far as possible derived from the photometric data studied
here. The bulk of datapoints in panels $a,b,c,e$ of
Fig.~\ref{definitions}, where the effect of reddening is dominant,
define clearly the color-excess ratios $E(r-i)/E(g-r)=0.48$,
$E(r-i)/E(i-J)=0.53$, $E(r-i)/E(i-H)=0.4$, and $E(r-i)/E(H-K)=3.1$,
respectively, in the direction of the Lagoon Nebula.
Moreover, the $(J-H,H-K)$ diagram (not shown) indicates
$E(J-H)/E(H-K)=1.9$. These relations are sufficient for the color-based
selections made in Sect.~\ref{members}, but not for estimates of optical
extinction or stellar masses from isochrones. Therefore, we need two
more relations in order to obtain a complete reddening law and
extinction estimates. These were chosen to be $A_J/A_V=0.282$ and
$A_K/A_V=0.175$ from Rieke and Lebofsky (1985), assuming that any
peculiar reddening law has negligible effect in the $J$ and $H$ bands.
}

\end{appendix}

\begin{acknowledgements}
We wish to thank an anonymous referee for his/her helpful suggestions.
Based on data products from observations made with ESO Telescopes at the
La Silla Paranal Observatory under programme ID 177.D-3023, as part of
the VST Photometric H$\alpha$ Survey of the Southern Galactic Plane and Bulge
(VPHAS+, www.vphas.eu).
This work has made use of data from the European Space Agency (ESA)
mission {\it Gaia} (\url{https://www.cosmos.esa.int/gaia}), processed
by the {\it Gaia} Data Processing and Analysis Consortium (DPAC,
\url{https://www.cosmos.esa.int/web/gaia/dpac/consortium}). Funding for
the DPAC has been provided by national institutions, in particular the
institutions participating in the {\it Gaia} Multilateral Agreement.
Also based on data products from VVV Survey observations made with the VISTA
telescope at the ESO Paranal Observatory under programme ID 179.B-2002.
The scientific results reported in this article are also based
on observations made by the Chandra X-ray Observatory.
This publication makes use of data products from the Two Micron All Sky
Survey, which is a joint project of the University of Massachusetts and
the Infrared Processing and Analysis Center/California Institute of
Technology, funded by the National Aeronautics and Space Administration
and the National Science Foundation.
This research makes use of the SIMBAD database and the Vizier catalog service,
operated at CDS, Strasbourg, France.
We also make heavy use of R: A language and environment for statistical
computing. R Foundation for Statistical Computing, Vienna, Austria.
(http://www.R-project.org/).
\end{acknowledgements}

\bibliographystyle{aa}

\begin{thebibliography}{}

\bibitem[Baraffe et al.(2015)]{2015A&A...577A..42B} Baraffe, I., Homeier,
D., Allard, F., \& Chabrier, G.\ 2015, \aap, 577, A42

\bibitem[Bailer-Jones et al.(2018)]{2018AJ....156...58B} Bailer-Jones, C.~A.~L., Rybizki, J., Fouesneau, M., Mantelet, G., \& Andrae, R.\ 2018, \aj, 156, 58

\bibitem[Chambers et al.(2016)]{2016arXiv161205560C} Chambers, K.~C.,
Magnier, E.~A., Metcalfe, N., et al.\ 2016, arXiv:1612.05560

\bibitem[Comeron \& Torra(1992)]{1992A&A...261...94C} Comeron, F., \& Torra, J.\ 1992, \aap, 261, 94

\bibitem[Comer{\'o}n(2001)]{2001ASPC..243..807C} Comer{\'o}n, F.\ 2001, From Darkness to Light: Origin and Evolution of Young Stellar Clusters, 243, 807

\bibitem[Damiani et al.(2004)]{2004ApJ...608..781D} Damiani, F.,
Flaccomio, E., Micela, G., et al.\ 2004, \apj, 608, 781

\bibitem[Damiani et al.(2006)]{2006A&A...459..477D} Damiani, F.,
Prisinzano, L., Micela, G., \& Sciortino, S.\ 2006a, \aap, 459, 477

\bibitem[Damiani et al.(2006)]{2006A&A...460..133D} Damiani, F., Micela,
G., Sciortino, S., et al.\ 2006b, \aap, 460, 133

\bibitem[Damiani et al.(2016)]{2016A&A...596A..82D} Damiani, F., Micela,
G., \& Sciortino, S.\ 2016, \aap, 596, A82

\bibitem[Damiani et al.(2017)]{2017A&A...604A.135D} Damiani, F., Bonito,
R., Prisinzano, L., et al.\ 2017a, \aap, 604, A135 

\bibitem[Damiani et al.(2017)]{2017A&A...602A.115D} Damiani, F.,
Pillitteri, I., \& Prisinzano, L.\ 2017b, \aap, 602, A115

\bibitem[Damiani(2018)]{2018arXiv180401905D} Damiani, F.\ 2018,
arXiv:1804.01905 (Paper~I)

\bibitem[Drew et al.(2014)]{2014MNRAS.440.2036D} Drew, J.~E.,
Gonzalez-Solares, E., Greimel, R., et al.\ 2014, \mnras, 440, 2036

\bibitem[Feigelson \& Montmerle(1999)]{1999ARA&A..37..363F} Feigelson, E.~D., \& Montmerle, T.\ 1999, \araa, 37, 363

\bibitem[Feigelson et al.(2013)]{2013ApJS..209...26F} Feigelson, E.~D.,
Townsley, L.~K., Broos, P.~S., et al.\ 2013, \apjs, 209, 26

\bibitem[Gaia Collaboration et al.(2016)]{2016A&A...595A...1G} Gaia
Collaboration, Prusti, T., de Bruijne, J.~H.~J., et al.\ 2016, \aap, 595, A1

\bibitem[Gaia Collaboration et al.(2018)]{2018arXiv180409365G} Gaia
Collaboration, Brown, A.~G.~A., Vallenari, A., et al.\ 2018,
arXiv:1804.09365

\bibitem[Guarcello et al.(2007)]{2007A&A...462..245G} Guarcello, M.~G.,
Prisinzano, L., Micela, G., et al.\ 2007, \aap, 462, 245

\bibitem[Haisch et al.(2001)]{2001ApJ...553L.153H} Haisch, K.~E., Jr., Lada, E.~A., \& Lada, C.~J.\ 2001, \apjl, 553, L153

\bibitem[Lawrence et al.(2007)]{2007MNRAS.379.1599L} Lawrence, A.,
Warren, S.~J., Almaini, O., et al.\ 2007, \mnras, 379, 1599

\bibitem[Kalari et al.(2015)]{2015MNRAS.453.1026K} Kalari, V.~M., Vink,
J.~S., Drew, J.~E., et al.\ 2015, \mnras, 453, 1026

\bibitem[King et al.(2013)]{2013ApJS..209...28K} King, R.~R., Naylor,
T., Broos, P.~S., Getman, K.~V., \& Feigelson, E.~D.\ 2013, \apjs, 209, 28

\bibitem[Kuhn et al.(2013)]{2013ApJS..209...27K} Kuhn, M.~A., Getman,
K.~V., Broos, P.~S., Townsley, L.~K., \& Feigelson, E.~D.\ 2013, \apjs,
209, 27

\bibitem[Kuhn et al.(2014)]{2014ApJ...787..107K} Kuhn, M.~A., Feigelson,
E.~D., Getman, K.~V., et al.\ 2014, \apj, 787, 107

\bibitem[Kuhn et al.(2015)]{2015ApJ...802...60K} Kuhn, M.~A., Getman,
K.~V., \& Feigelson, E.~D.\ 2015, \apj, 802, 60

\bibitem[Kuhn et al.(2018)]{2018arXiv180702115K} Kuhn, M.~A., Hillenbrand, L.~A., Sills, A., Feigelson, E.~D., \& Getman, K.~V.\ 2018, arXiv:1807.02115

\bibitem[Kumar \& Anandarao(2010)]{2010MNRAS.407.1170K} Kumar, D.~L.,
\& Anandarao, B.~G.\ 2010, \mnras, 407, 1170

\bibitem[Luri et al.(2018)]{2018arXiv180409376L} Luri, X., Brown,
A.~G.~A., Sarro, L.~M., et al.\ 2018, arXiv:1804.09376

\bibitem[Minniti et al.(2011)]{2011BAAA...54..265M} Minniti, D.,
Clari{\'a}, J.~J., Saito, R.~K., et al.\ 2011, Boletin de la Asociacion
Argentina de Astronomia La Plata Argentina, 54, 265

\bibitem[Prisinzano et al.(2005)]{pris04} Prisinzano, L., Damiani, F.,
Micela, G., Sciortino, S. 2005, \aap, 430, 941

\bibitem[Prisinzano et al.(2007)]{2007A&A...462..123P} Prisinzano, L.,
Damiani, F., Micela, G., \& Pillitteri, I.\ 2007, \aap, 462, 123

\bibitem[Prisinzano et al.(2012)]{2012A&A...546A...9P} Prisinzano, L.,
Micela, G., Sciortino, S., Affer, L., \& Damiani, F.\ 2012, \aap, 546, A9

\bibitem[Prisinzano et al.(2018)]{2018arXiv180606625P} Prisinzano, L.,
Damiani, F., Guarcello, M.~G., et al.\ 2018, arXiv:1806.06625

\bibitem[Siess et al.(2000)]{2000A&A...358..593S} Siess, L., Dufour, E.,
\& Forestini, M.\ 2000, \aap, 358, 593

\bibitem[Sung et al.(2000)]{2000AJ....120..333S} Sung, H., Chun, M.-Y.,
\& Bessell, M.~S.\ 2000, \aj, 120, 333

\bibitem[Tenorio-Tagle et al.(1987)]{1987A&A...179..219T} Tenorio-Tagle, G., Franco, J., Bodenheimer, P., \& Rozyczka, M.\ 1987, \aap, 179, 219

\bibitem[Tothill et al.(2008)]{2008hsf2.book..533T} Tothill, N.~F.~H.,
Gagn{\'e}, M., Stecklum, B., \& Kenworthy, M.~A.\ 2008, Handbook of Star
Forming Regions, Volume II, 5, 533

\bibitem[van den Ancker et al.(1997)]{1997A&AS..123...63V} van den Ancker, M.~E., The, P.~S., Feinstein, A., et al.\ 1997, \aaps, 123, 63

\bibitem[Venuti et al.(2018)]{2018arXiv181102731V} Venuti, L., Damiani, F., \& Prisinzano, L.\ 2018, arXiv:1811.02731

\bibitem[Wright et al.(1977)]{1977AJ.....82..132W} Wright, E.~L., Lada,
C.~J., Fazio, G.~G., Low, F.~J., \& Kleinmann, D.~E.\ 1977, \aj, 82, 132

\end{thebibliography}

\begin{landscape}
\begin{table}
\centering
\caption{Optical/NIR photometry for all candidate members of NGC~6530. Full table in electronic format only.
\label{table-members}}
\begin{tabular}{rccccccccccccccccc}
  \hline
Seq & RA & Dec & VPHAS Id & $i$ & $r-i$ & $r-H\alpha$ & $g-r$ & $u-g$ & $J$ & $J-H$ & $H-K$ & IR & \ha\ & UV & M & X-ray & Gaia \\
no.\ & (J2000) & (J2000) & & & & & & & & & & exc.\ & exc.\ & exc.\ & star & det.\ &  \\
  \hline
1 & 270.75503 & -24.47542 & J180301.2-242831.4 & 20.95 & 1.80 &  &  &  & 17.14 & 0.75 & 0.27 &  &  &  & Y &  &  \\ 
  2 & 270.77272 & -24.45467 & J180305.5-242716.8 & 20.45 & 1.73 &  &  &  & 16.78 & 0.91 & 0.35 &  &  &  & Y &  &  \\ 
  3 & 270.86873 & -24.45169 & J180328.5-242706.1 &  &  &  &  &  &  &  &  &  &  &  &  &  & Y \\ 
  4 & 270.82115 & -24.49788 & J180317.1-242952.3 &  &  &  &  &  &  &  &  &  &  &  &  &  & Y \\ 
  5 & 270.85809 & -24.46839 & J180325.9-242806.2 &  &  &  &  & 0.94 &  &  &  &  &  &  &  & Y &  \\ 
  6 & 270.82926 & -24.46857 & J180319.0-242806.8 &  &  &  &  &  &  &  &  &  &  &  &  & Y &  \\ 
  7 & 270.76139 & -24.52889 & J180302.7-243143.9 & 20.11 & 1.48 &  &  &  & 14.52 & 1.75 & 0.81 & Y &  &  &  &  &  \\ 
  8 & 270.82775 & -24.42138 & J180318.7-242516.9 & 20.77 &  &  &  &  & 17.32 & 0.80 & 0.22 &  &  &  &  & Y &  \\ 
  9 & 270.81253 & -24.47325 & J180315.0-242823.6 & 19.61 & 1.66 &  & 1.65 &  &  &  &  &  &  &  & Y &  &  \\ 
  10 & 270.73951 & -24.54836 & J180257.5-243254.0 & 19.85 & 1.49 & 0.67 & 1.83 &  &  &  &  &  &  &  & Y &  &  \\ 
  11 & 270.76578 & -24.51855 & J180303.8-243106.7 & 18.25 & 1.02 & 0.59 & 1.81 &  &  &  &  &  &  &  &  &  & Y \\ 
  12 & 270.72789 & -24.50358 & J180254.7-243012.8 & 18.10 &  &  & 1.66 &  &  &  &  &  &  &  &  &  & Y \\ 
  13 & 270.77432 & -24.52539 & J180305.8-243131.3 & 20.81 & 1.73 &  &  &  & 17.43 & 0.72 & 0.17 &  &  &  & Y &  &  \\ 
  14 & 270.77804 & -24.49452 & J180306.7-242940.2 & 18.39 &  & 0.51 & 2.22 &  & 14.67 & 0.82 & 0.33 &  &  &  &  &  & Y \\ 
  15 & 270.81611 & -24.50425 & J180315.9-243015.2 & 16.06 &  &  & 1.49 & 1.39 &  &  &  &  &  &  &  &  & Y \\ 
  16 & 270.76885 & -24.41684 & J180304.5-242500.6 & 20.20 & 1.68 &  &  &  & 16.51 & 0.96 & 0.36 &  &  &  & Y &  &  \\ 
  17 & 270.84740 & -24.48966 & J180323.4-242922.7 &  &  &  &  &  & 14.48 & 0.98 & 0.43 &  &  &  &  &  & Y \\ 
  18 & 270.58659 & -24.52602 & J180220.8-243133.7 & 20.27 & 1.87 & 0.94 &  &  & 16.27 & 0.90 & 0.39 &  &  &  & Y &  &  \\ 
  19 & 270.83647 & -24.44672 & J180320.8-242648.2 & 20.49 & 1.94 &  &  &  & 15.07 & 1.36 & 0.56 &  &  &  &  & Y &  \\ 
  20 & 270.77585 & -24.48878 & J180306.2-242919.5 &  &  &  &  &  & 13.82 & 0.70 & 0.31 &  &  &  &  &  & Y \\ 
   \hline
\end{tabular}
\end{table}
\end{landscape}

\end{document}